\documentclass{aa}  
\usepackage{natbib}
\bibpunct{(}{)}{;}{a}{}{,} 
\usepackage{graphicx}
\usepackage{txfonts}

\newcommand{\cl}{Cl~0332-2742} 
\begin{document}

\title{GMASS ultradeep spectroscopy of galaxies at $z \sim
    2$\footnote{Based on observations of the VLT Large Program
    173.A-0687 carried out at the European Southern Observatory,
    Paranal, Chile.} }

   \subtitle{V. Witnessing the assembly at $z = 1.6$ of a galaxy cluster}

   \author{J. Kurk\inst{1,2} \and
          A. Cimatti\inst{3} \and
          G. Zamorani\inst{4} \and
          C. Halliday\inst{2} \and
          M. Mignoli\inst{4} \and
          L. Pozzetti\inst{4} \and
          E. Daddi\inst{5} \and
          P. Rosati\inst{6} \and
          M. Dickinson\inst{7} \and
          M. Bolzonella\inst{4} \and
          P. Cassata\inst{8} \and
          A. Renzini\inst{9} \and
          A. Franceschini\inst{10} \and
          G. Rodighiero\inst{10} \and
          S. Berta\inst{11}
          }

   \offprints{J. Kurk}

   \institute{Max-Planck-Institut f\"ur Astronomie, K\"onigstuhl 17, 
              D-69117, Heidelberg\\
              \email{kurk@mpia.de}
         \and
             INAF-Osservatorio Astrofisico di Arcetri, 
               Largo E. Fermi 5, I-50125, Firenze\\
         \and
             Universit\`a di Bologna, Dipartimento di Astronomia, 
               Via Ranzani 1, I-40127, Bologna\\
         \and
             INAF-Osservatorio Astronomico di Bologna,
               Via Ranzani 1, I-40127, Bologna\\
         \and
             CEA, Laboratoire AIM - CNRS - Universit\'e Paris Diderot, 
                Irfu/SAp, Orme des Merisiers, F-91191 Gif-sur-Yvette, France\\
         \and
             European Southern Observatory, 
               Karl-Schwarzschild-Strasse 2, D-85748, Garching bei M\"unchen\\
         \and
             NOAO-Tucson, 950 North Cherry Avenue, Tucson, AZ 85719, USA\\
         \and
             Department of Astronomy, University of Massachusetts, 
             710 North Pleasant Street, Amherst, MA 01003, USA\\
         \and
             INAF-Osservatorio Astronomico di Padova,
               Vicolo dell'Osservatorio 5, I-35122, Padova\\
         \and
             Universit\`a di Padova, Dipartimento di Astronomia, 
               Vicolo dell'Osservatorio 2, I-35122, Padova\\
         \and
             Max-Planck-Institut f\"ur extraterrestrische Physik,
             Giessenbachstra\ss{}e, D-85748, Garching bei M\"unchen
             }

   \date{Received ; accepted }

 
  \abstract
  {Clusters of galaxies represent important laboratories for studying
  galaxy evolution and formation. Well established and relaxed
  clusters are known below $z < 1.4$, as well as, clusters in
  formation found close to radio galaxies at $z > 2$, but the
  in-between redshift range, during which clusters are expected to
  undergo significant changes, is almost unexplored.}
  {By studying a galaxy overdensity in redshift and angular
  distribution at $z = 1.6$, uncovered in the Galaxy Mass Assembly
  ultra-deep Spectroscopic Survey (GMASS), we provide insight into the
  evolution of cluster galaxies at high redshift.}
  {We present a study of the significance of the galaxy overdensity at
  $z = 1.6$, \cl, its velocity dispersion, and X-ray emission.
  We reveal colour bimodality for cluster members and compare the
  properties of members of \cl\ with galaxies outside the
  overdensity.}
{From the redshifts of the 42 overdensity members, we find a velocity
  dispersion of 500 km\,s$^{-1}$.  We conservatively estimate the
  overdensity in redshift space for the spike at $z=1.6$ in the GMASS
  to be 8.3$\pm$1.5.  A map of the surface density of galaxies at
  $z=1.6$ in the GMASS field shows that its structure is irregular
  with several filaments and local overdensities. The differences in
  the physical properties of \cl\ member and field galaxies agree with
  the latest hierarchical galaxy formation models: for overdensity
  members, the star formation rate (SFR), and specific SFR, is
  approximately 50\% lower than for the field galaxies; overdensity
  galaxies are twice the age, on average, of field galaxies; and there
  is a higher proportion of both massive ($M > 10^{10.7}M_\odot$), and
  early--type galaxies, inside \cl\ than in the field.  Among the
    42 members, seven have spectra consistent with being passively evolving,
    massive galaxies.  These are all located within an area where the
    number density of $z=1.6$ galaxies is highest.  In a $z-J$
  colour-magnitude diagram, the photometric data of these early-type
  galaxies are in close agreement with a theoretical red sequence of a
  galaxy cluster at redshift $z = 1.6$, which formed most of its stars
  in a short burst of star formation at $z \sim 3$.}
{We conclude that the redshift spike at $z=1.6$ in the GMASS
    field represents a sheet--like structure in the cosmic web and the
    area with the highest surface density within this structure,
    containing already seven passively evolving galaxies, will evolve
    into a cluster of galaxies at a later time.}


   \maketitle
%

\section{Introduction}\label{sec:introduction}

Clusters of galaxies are the most massive structures in the
  universe.  Because the evolution of structure in the early universe
  depends on mass, galaxy clusters play an important role in cosmology
  and can help to constrain cosmological parameters
  \citep[e.g.,][]{rap08}.  Clusters are also powerful tools for studying
  galaxy evolution.

It is well known that environment plays an important role in galaxy
evolution.  Correlations between the environment of galaxies,
  e.g., in terms of galaxy density, and several galaxy properties,
  such as the age and metallicity of their stellar populations, star
  formation rates, and morphologies have been studied, in individual
  galaxy clusters and also as a function of redshift.  Early
observations of cluster galaxies out to $z = 0.5$ by, for example,
\citet{but78,but84} showed little evidence of evolution in the colours
of the red cluster population, but detected an increase in the
fraction of blue galaxies, which appeared to avoid the cluster
centre. Although this \emph{Butcher-Oemler} effect has been disputed,
the general conclusion still stands \citep[e.g.,][]{ell01, tan05,
  and06, tra07, fas08}.  \citet{bal07} find that, for a given
  stellar mass and redshift, fewer galaxies in groups show strong
  [\ion{O}{II}] emission than in field galaxies.  The evolution of
  star formation rate appears to depend on both galaxy mass and galaxy
  environment \citep{bal06}.  The observations by \citeauthor{bal06}
  of a large sample of galaxies at $z<0.1$ were compared directly with
  the semi-analytic models of galaxy formation by \citet{bow06} and
  \citet{cro06}, showing closer agreement with the former.  This
  colour-density relation has also been confirmed at higher redshifts,
  up to $z \sim 1.5$ \citep{cuc06,coo07}.  At these high redshifts,
  few massive systems are known, as \citet{pog08} note.  In addition,
  distant cluster surveys have studied primarily massive clusters.
  \citeauthor{pog08}, therefore, employ data from the ESO Distant
  Cluster Survey to investigate the relation between star formation
  activity, morphology and local galaxy density in galaxy groups and
  clusters at $z = 0.4 - 0.8$.  They detect proportionally fewer
  galaxies with ongoing star formation in regions of higher projected
  density in both distant and nearby clusters, but no evidence that
  the star formation properties of a given Hubble type or galaxy mass
  depend on local environment.

  A striking feature in the spectrum of an old stellar population is
  the 4000\,\AA\ break, above which the light of old (red) stars
  dominate over the rare young stars.  This feature causes the
  occurrence of a sequence of red (passive) galaxies in a (suitable)
  colour--magnitude diagram of galaxy clusters.  The tight sequence
  observed in both nearby and distant clusters (up to $z=1.24$)
  suggests that the stars in these galaxies formed at very high
  redshift in a relatively short time \citep[e.g.,][]{sta98,bla03}.
  An analysis of stellar population properties and morphologies of
  red--sequence galaxies in clusters and groups from $z\sim0.75$ to
  $z\sim0.45$ by \citet{san09} shows that the rate at which
  red--sequence galaxies evolve depends on their mass, as less massive
  galaxies show evidence for a more extended star formation history
  than more massive galaxies do.  One therefore expects that, at
  higher redshift, the scatter in colour increases as the epoch when
  the stellar populations were formed is approached.  Indeed, changes
  in the red sequence in galaxy structures have been observed between
  low redshift and redshifts $z=2.2 - 3.1$ \citep{kod07,zir08}, in the
  sense that the faint end of the red sequence becomes less populated,
  although a dependence on cluster richness may also play a role
  \citep{and08}.

Distant groups and clusters of galaxies, therefore, provide important
environments for the study of galaxy evolution, especially if both
passive and actively star--forming galaxy populations can be studied
in detail. For the overdensities of galaxies known at high redshift,
this is often problematic because their cluster members are selected
on the basis of either their star formation activity \citep[blue
members, e.g.,][]{kur00, pen00, ven02, ste05}, or lack of it
\citep[red members, e.g.,][]{bes03, kod07, mcc07}. The most well studied
high-redshift systems confirmed to be clusters on the basis of their
X-ray emission, are found between $z = 1.0$ and $z = 1.4$, redshifts
for which spectroscopic data of both the red and blue populations can
be practically acquired \citep[e.g., at $z=1.26$, ][]{dem07, mei06}. These
clusters were discovered either by their extended X-ray emission due
to hot cluster gas \citep[RX J0848.9+4452 at $z=1.26$, XMMU
J2235.3-2557 at $z=1.39$, XMMXCS J2215.9-1738 at $z = 1.45$,
][ respectively]{ros99, mul05, sta06}, or by studying near-infrared
(NIR) photometry of red cluster members \citep[ISCS J143809+341419 at
$z = 1.41$][]{sta05}.

We present a galaxy overdensity at $z = 1.61$, possibly a cluster
  under assembly, containing both red and blue galaxy populations.  In
  Sect.~\ref{sec:thedata}, we introduce the overdensity as described
  in other papers and describe the data set that we used to gather
  additional data.  We also discuss the selection bias that may be
  present in our data.  In Sect.~\ref{sec:overdensityproperties}, we
  describe the properties of the overdensity, including its velocity
  dispersion, an estimate of its mass, and the presence of a red
  sequence of galaxies.  In Sect.~\ref{sec:galaxyproperties}, the
  properties of the individual galaxies in the overdensity are
  described, in relation also to the local density, and in comparison
  with galaxies in a field sample.  The overdensity and its galaxies
  are discussed in the context of the literature in
  Sect.~\ref{sec:discussion}. In Sect.~\ref{sec:summary}, the paper
  is summarized.

The following cosmological parameters are used throughout this paper:
$H_0 = 70$ km s$^{-1}$ Mpc$^{-1}$, $\Omega_{\rm m} = 0.3, \Omega_{\rm
\Lambda} = 0.7, \Omega_{\rm k} = 0$. Magnitudes are quoted in the AB
system.


\section{The data}\label{sec:thedata}

\subsection{An overdensity at $z=1.6$}\label{ssec:peak_intro}

The overdensity at $z=1.6$ was detected by means of its clustering in
the redshift space of the deep spectroscopic GMASS survey (Kurk et
al., in preparation). We refer to this overdensity as \cl, which
corresponds to the coordinates of the galaxy at the centre of the
number density isosurfaces (see below). The GMASS survey was carried
out in the field of GOODS-S, where several overdensities at various
redshifts have been detected. \citet{gil03} described five spikes in
the X-ray source redshift distribution of this field; the second
highest in terms of redshift occurs at $z=1.62$ and contains five
sources.  The Poissonian probability of observing such a high number
of sources, given the background count rate, is only $4\times10^{-3}$,
which implies that this aggregation represents a significant
group. \citet{van06} also described spikes in the redshift
distribution using data from the ESO/GOODS spectroscopy program of
faint galaxies in the GOODS-S field. They reported that 20 galaxies
have redshifts of $z \sim 1.61$ and eight other galaxies had redshift
data from other surveys; all of these galaxies are distributed
spatially in an apparently non-uniform way. The probability of
detecting such a significant peak in redshift space by chance is less
than $7 \times 10^{-5}$, according to Monte Carlo simulations of the
smoothed redshift distribution of the 501 confirmed redshifts, after
the second stage of the ESO/GOODS spectroscopy program.

Three old, fully-assembled, massive ($> 10^{11} $M$_\odot$) spheroidal
galaxies within the spike at $z=1.61$ were detected in the K20 survey
\citep{cim02} and described in \citet{cim04}. In another article based
on GMASS results, \citet{cim08} investigated the properties of
thirteen old, passive galaxies at $z > 1.4$, seven of which are at $z
= 1.6$.  These galaxies have stellar-mass surface-densities that are
$\sim 1$ dex higher than local spheroids of similar mass.

An overdensity at $z=1.6$ was also discerned in photometric redshift
space by \citet{cas07}; these authors applied their own (2+1)D cluster
finding algorithm \citep[see also][]{tre07} to the GOODS-South Field
and discovered a localized overdensity, embedded in a larger scale
overdensity of galaxies at $z = 1.6$. Their method was based on an
estimate of three-dimensional densities and considered simultaneously
angular positions and photometric redshifts (in this case from the
GOODS-MUSIC catalogue by \citealt{gra06}). From their analysis,
several major peaks at different redshifts emerged, the peak at
$z \sim 1.6$ being the most significant at $z > 1$ in the GOODS-South
field.  Within this large--scale overdensity \citep[for which
additional spectroscopic evidence was presented by][]{cim02, cim04,
  gil03, van05, van06}, they isolated a compact, higher density peak,
centred approximately on RA=03$^h$32$^m$29.98$^s$,
DEC=-27$^\circ$42\arcmin35.99\arcsec, which was identified to be a
cluster of an approximate total extension of $3 \times 3$ Mpc
(comoving), centred on a large star-forming, remarkably irregular
galaxy (see Fig.\ 2 in \citealt{cas07}). In this paper, we define a
circular \emph{high--density} region, centred on the same coordinates,
of radius 1 Mpc (physical). The reasons for this choice are
explained in Sect.~\ref{sec:angular}. On the basis of the irregular
morphology of the galaxy overdensity, its low total X-ray emission,
and its estimated mass of (1.4 -- 4.1) $\times 10^{14}$ M$_\odot$,
\citeauthor{cas07} concluded that this group/poor cluster had not yet
reached its virial equilibrium.

\subsection{The GMASS sample}

GMASS ({\it ``Galaxy Mass Assembly ultra-deep Spectroscopic Survey''}
\footnote{http://www.arcetri.astro.it/$\sim$cimatti/gmass/gmass.html})
is a project based on an ESO VLT Large Program (173.A-0687, P.I.\
A.~Cimatti).  The GMASS project and sample are described in detail by
J.\ D.\ Kurk et al.\ (2009, in preparation, hereafter Paper VI), and
we recall only its main features. The GMASS sample was selected from a
region of $6.8\times6.8$ arcmin$^2$, located in the GOODS-South
field\footnote{http://www.stsci.edu/science/goods} \citep[][ Dickinson
et al., in preparation]{gia04}; all sources detected in the publicly
available $4.5\,\mu$m image, acquired using IRAC on the {\sl Spitzer
  Space Telescope}, were extracted to a limiting magnitude of
$m_{4.5}<23.0$ (2.3\,$\mu$Jy), which provided a selection related
closely to stellar mass. The GMASS sample includes 1277 unblended
objects to $m_{4.5}<23.0$, which have photometry from the NUV to MIR
and SED fits obtained using \citet[][ M05]{mar05} templates providing
stellar masses, star formation rates (SFRs), and other galaxy
properties \citep[see Sec.~\ref{sec:galaxyproperties} and][]{poz07}.

The GMASS ESO VLT+FORS2 optical spectroscopy was focused on galaxies
pre--selected using a cut in photometric redshift of $z_{\rm
  phot}>1.4$ and two cuts in the optical magnitudes ($B<26.5$, $I<
26.5$). This selection produced 221 spectroscopic targets, 174 of
which were actually observed (called the spectroscopic sample,
hereafter). The integration times were extremely long (up to 32 hours
per mask and some targets were included in multiple masks), and the
utility of the spectroscopic data was optimized by acquiring spectra
at either blue (4000--6000 \AA) or red (6000--10000 \AA) wavelengths,
depending on the colours and photometric SEDs of the targets, called
the blue and red masks, respectively, hereafter. Despite the faintness
of the targets, the GMASS spectroscopy program was very successful:
we were able to determine redshifts for 150 out of the 174
  targeted sources (86\%), 132 of which turned out to have $z_{\rm
    spec}>1.4$.  This implies that fractions of $\geq$76\% and 88\% of
  all targeted sources and those with successfully determined
  redshifts, respectively, are at $z_{\rm spec} >1.4$.  In addition to
  the 18 objects from the spectroscopic sample with $z_{\rm
    spec}<1.4$, filler objects used to complete the masks delivered
  another 27 redshifts of $z_{\rm spec}<1.4$ and three of $z_{\rm
    spec}>1.4$. 

\begin{figure}
  \centering
  \includegraphics[width=\columnwidth]{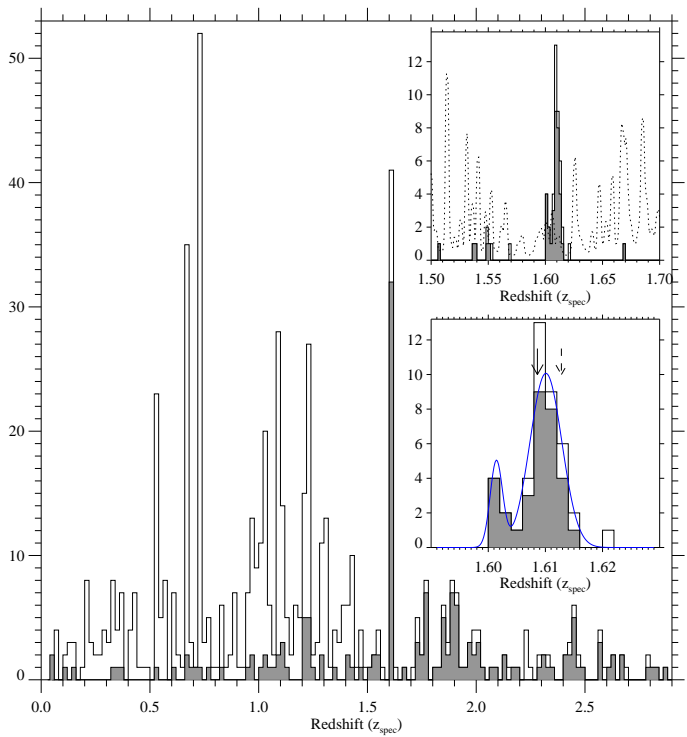}
  \caption{ Redshift distribution of spectroscopic redshifts of
    objects in the GMASS field displayed with bin sizes of $\Delta z =
    0.02$ (main panel) and 0.002 (insets). Redshifts determined within
    the GMASS survey are indicated by the grey histogram. Within the
    upper inset, the sky background relevant to [\ion{O}{II}]
    detection is indicated by a dashed line. Within the lower inset,
    which has the same bin size but is zoomed in on a smaller redshift
    range, including the spike only, a curve with two Gaussian
    functions fitted to the histogram is overlayed. In addition, a
    dashed histogram showing the 21 galaxies within the high density
    region (see text) is included. The solid arrow indicates the
    redshift of the brightest member galaxy, while the dashed arrow
    indicates the redshift of the central (see text) galaxy.}
  \label{fig:zdistr_dz0.02}
\end{figure}

Figure~\ref{fig:zdistr_dz0.02} shows the distribution of all known
spectroscopic redshifts (including those of other surveys in GOODS-S,
see Paper VI) of galaxies within the GMASS catalogue (stars are omitted),
up to $z = 2.9$. Several peaks in the distribution are prominent, the
peak at $z = 1.6$ being the most significant at $z > 1.4$. Although
this peak was previously known to exist (see Sect.~\ref{ssec:peak_intro} for
references), the availability of 136 spectroscopic redshifts at $z >
1.4$ from GMASS allows a far more careful study than possible
before. The redshift peak at $z=1.6$ may indicate the presence of a
galaxy cluster or a more extended galaxy overdensity (such as a
sheet). Because the dynamical state and richness of \cl\ is a priori
unknown, we refer to the galaxy aggregation as a galaxy group or cluster
in formation and to the galaxies considered to be part of the spike in
the redshift distribution as \emph{members} or member galaxies.

\subsection{Assessment of the spectroscopic selection bias}
\label{sec:spectroscopic_bias}

Our criteria for the selection of the spectroscopic sample may create
a selection bias, which should have a minor influence on the redshift
distribution in the range $1.4 < z < 1.8$ considered here. A bias
introduced by the success of the redshift determination, could be more
important. At wavelengths for which [\ion{O}{II}] is detected, in
galaxies at $z \sim 1.6$, several sky lines of a range of intensities,
in the observed frame, are present. These lines can, in principle, have
a serious effect on the measurement of emission-line wavelengths, and
spurious peaks in the redshift distribution can be produced. We find,
however, that this effect has a minor influence on our redshift
determination.

In Sect.~\ref{ssec:velocitydispersion}, we show that the 42 galaxies
in the redshift range $1.600 < z_{\rm spec} < 1.622$ are likely to be
members of the galaxy overdensity.  Of these, 27 have redshifts
determined only by GMASS, while ten have redshifts determined only by
ESO/GOODS spectroscopy \citep{van06}. The remaining five have
redshifts determined by GMASS spectroscopy, but four of these were
also observed by \citeauthor{van06} (two with an uncertain redshift,
one of them being inconsistent with the more accurate GMASS redshift),
and one was part of the K20 survey (but with an uncertain and
inconsistent redshift). We examined all 42 spectra in detail and
assessed whether redshift determination would have been possible
without the detection of the [\ion{O}{II}] line, which was the case
for 20 (i.e., 48\%) of the galaxy spectra with strong UV absorption
lines. For eight of these spectra, the region about $\lambda =
9690$\,\AA, which corresponds to the wavelength of the [\ion{O}{II}]
line at $z=1.61$, was not observed at all, while for an additional
three the region was observed, but the [\ion{O}{II}] line was not
detected.  A wavelength--dependent selection function is not an issue
for these galaxy spectra, because the main UV absorption lines are
observed in an observed-frame wavelength range for which no strong sky
lines are detected.

As a second check, we assessed precisely the sky line situation about
9690\,\AA.  The result is shown in the upper inset of
Fig.~\ref{fig:zdistr_dz0.02}, where we have translated the wavelengths
of the sky spectrum to the redshift relevant to [\ion{O}{II}]
detection.  We note that the region corresponding to $1.600 < z <
1.616$ is populated by weak sky lines.  In contrast, the wavelength
range corresponding to $1.570 < z < 1.600$, for which not a single
redshift was determined, has far lower sky line contamination. In
addition, we note that in the redshift range $1.616 < z < 1.622$,
corresponding to an observed wavelength range for which no sky lines
are present, only one galaxy is detected.

We conclude that the observed-frame wavelength range affected by mild
sky-line contamination is significantly wider than the range
encompassing the overdensity redshift peak. Sky lines have, therefore,
a minor influence, if any, on the measurement of redshift for the
twelve members with GMASS spectroscopy, for which redshift was
measured using [\ion{O}{II}], and the ten members for which redshifts
were measured by the ESO/GOODS survey.

We finally note that the redshift peak is also present in the
photometric redshift distribution (see Fig.\ 1 in paper VI), which is
subject to different selection biases and therefore unlikely to
generate a peak at the same redshift by chance.

\section{Overdensity properties}\label{sec:overdensityproperties}

\subsection{Angular distribution}
\label{sec:angular}

\begin{figure*}
  \centering
  \includegraphics[width=0.9\linewidth,clip=]{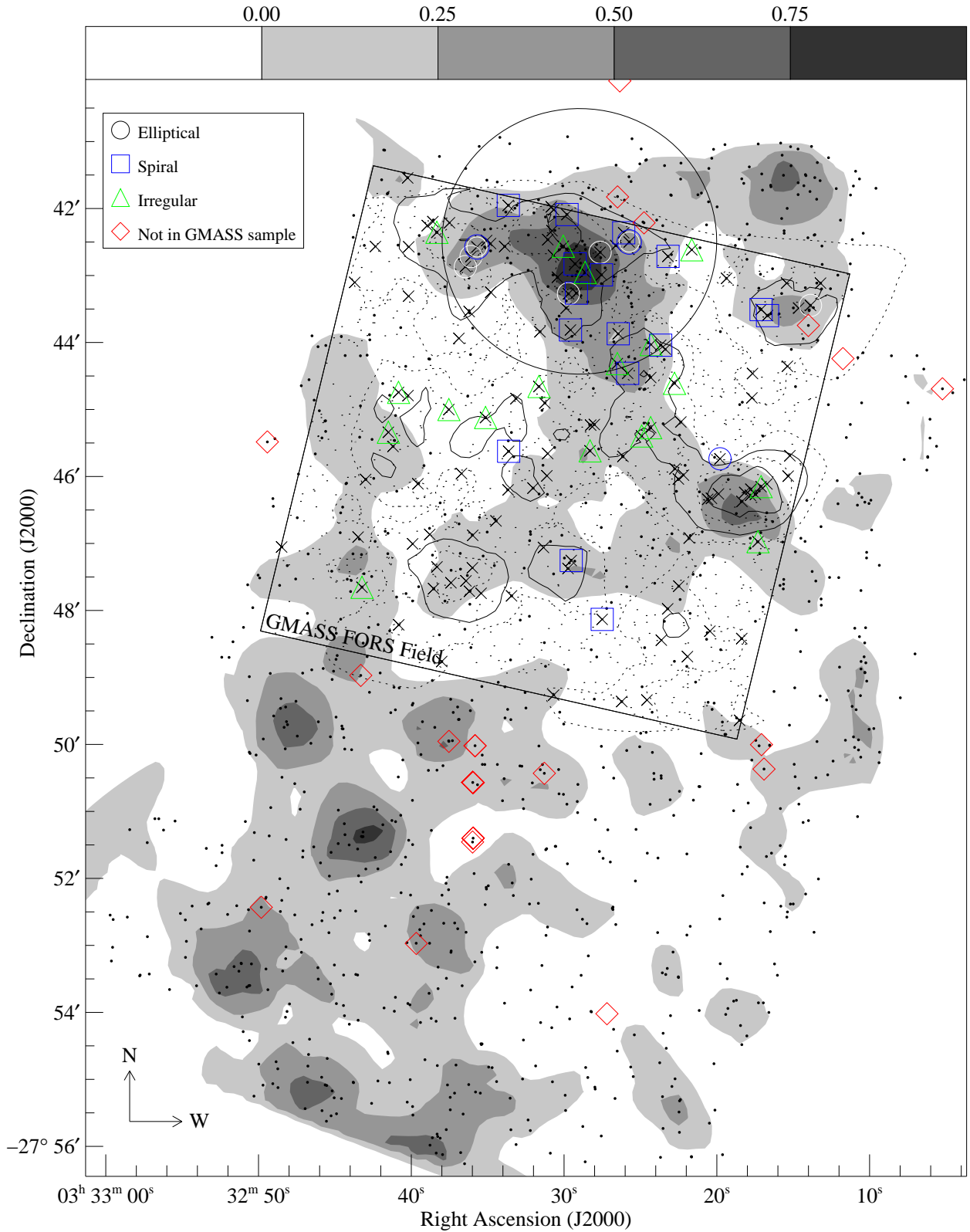}
  \caption{Density maps of galaxies at $z = 1.6$ in the GOODS-South
    field.  Filled (solid and dashed) contours indicate the density of
    $1.43 < z_{\rm phot} < 1.77$ galaxies based on the GOODS-MUSIC
    (GMASS) catalogue photometric (and spectroscopic) redshifts.
    Contours indicate 0.0, 0.25, 0.5, and 0.75$\times$ the maximum
    overdensity above the median density of $z=1.6$ galaxies.  This
    maximum overdensity is 1.9 (4.6) times the median density.  Large
    dots (crosses) indicate the positions of GOODS-MUSIC (GMASS)
    galaxies in the redshift range.  Small dots indicate galaxies in
    the GMASS catalogue outside the redshift range.  \cl\ galaxies with
    confirmed redshifts are indicated by additional symbols according
    to their morphology: circles for elliptical, squares for spiral,
    and triangles for irregular galaxies.  The elliptical galaxies
    have early--type spectra, except for the two indicated with double
    circles, which have intermediate--type spectra.  All others have
    late--type spectra.  Diamonds indicate galaxies with redshifts
    confirmed by ESO GOODS spectroscopy to be in the range $1.600 <
    z_{\rm spec} < 1.622$, but are not in the GMASS sample. }
  \label{fig:geom}
\end{figure*}

\begin{figure*}
  \centering
  \includegraphics[width=\linewidth,clip=]{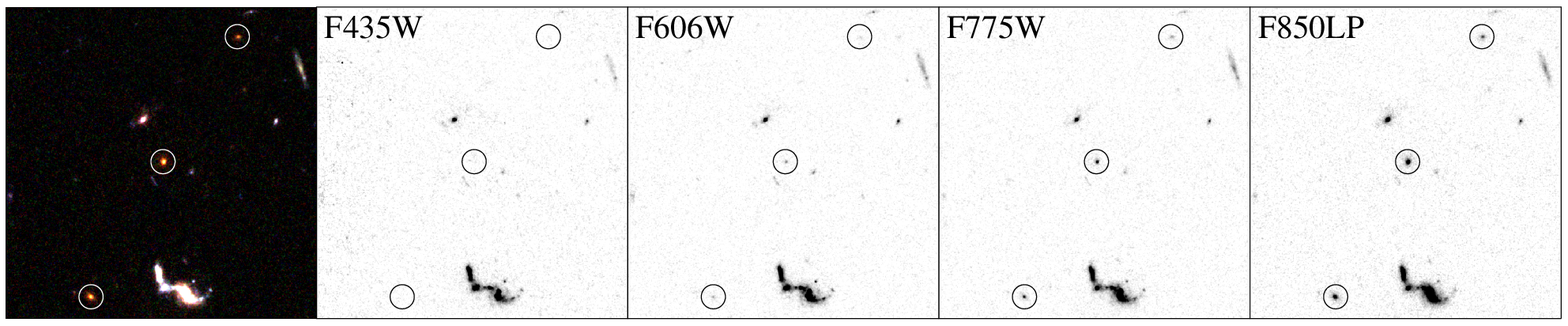}
  \caption{The 20\arcsec\ long triplet of $z=1.61$ early--type
    galaxies (as indicated by circles).  Shown are a three--colour
    (F606W, F775W and F850LP) image and the individual ACS bands,
    provided by the publicly available HST GOODS observations.  The
    colour scaling covers the range from 0 to 25$\sigma$, and 0 to
    40$\sigma$, respectively, where $\sigma$ is the dispersion around
    the mean background level (which is zero by default here).  The
    images are 21\arcsec\ on a side.  Note the much bluer foreground
    (and possibly background) galaxies.}
  \label{fig:triplet}
\end{figure*}

Figure~\ref{fig:geom} shows the geometry of the GMASS field,
indicating objects with known spectroscopic redshifts by small dots,
galaxy members (i.e., objects with $1.600 < z_{\rm spec} < 1.622$) by
crosses or diamonds, and photometric redshifts in the range $1.43 <
z_{\rm phot} < 1.77$ by large dots. There is a distinct concentration
of galaxy members towards the northern edge of the GMASS field. In
addition, among the galaxy members, the red, early, and
intermediate-type galaxies are most strongly concentrated towards the
north. These types were determined from the spectra of the galaxies on
the following basis: class 1 (early-type) have absorption lines only,
class 2 (late-type) have strong emission lines, and class 1.5
(intermediate-type) have absorption lines but some emission lines
(most commonly [\ion{O}{II}]) are also present.

The brightest confirmed galaxy member (GMASS\,2148 at $z = 1.609$,
close to the mean redshift of the spike) is the central galaxy within
a striking linear triplet of early-type galaxies, having a length of
20\arcsec\ (or 169\,kpc) in the north-east of the GMASS field (see
Fig.~\ref{fig:triplet}). The other two members of the triplet
(GMASS\,2196 and 2543) have redshifts $z = 1.614$ and 1.612
(respectively) and the largest difference in redshift corresponds to
$\sim 575$ km s$^{-1}$.  The crossing time of a galaxy in this triplet
is $\sim 3 \times 10^8$ yr, indicating that these systems are possibly
in the process of merging, in which case they would form one central
massive galaxy with a mass of $3 \times 10^{11} M_\odot$.  In a
colour-magnitude diagram of this field, GMASS\,2196 and 2543 are also
located very close to GMASS\,2148, the latter forming the tip of the
red sequence (see Sect.~\ref{ssec:cm}). It is remarkable that this
triplet of red galaxies lies about 1.5\arcmin\ (i.e., 760\,kpc)
eastward of the centre of the high density region and X-ray emission
described by \citet{cas07}. This may be an indication that the
cluster is not yet relaxed.

We also show in Fig.~\ref{fig:geom} the density of $z \sim 1.6$
galaxies based on photometric redshifts in the GOODS-MUSIC catalogue
as filled contours \citep[this density map is comparable to that
presented by][]{cas07}. This density is computed by counting the
number of neighbouring galaxies at $z \sim 1.6$ within a radius
(projected on the sky) of about 300 (physical) kpc (roughly the scale
of galaxy groups). To normalize this number and thereby correct for
edge effects and the presence of bright stars, this value is divided
by the total number of galaxies in the catalogue within this
radius. Contours are subsequently drawn using triangulation to obtain
a regular grid. This process involves some extrapolation at the edges
of the grid that produces obvious artifacts in the density maps, which
should be disregarded by the reader. In Fig.~\ref{fig:geom}, the
redshift interval of members is taken to be $1.43 < z_{\rm phot} <
1.77$, which is based on a comparison of photometric redshifts within
the MUSIC catalogue with the redshifts of spectroscopically--confirmed
members. The solid and dashed contours are based on the photometric
redshifts in the GMASS catalogue, using the same redshift interval as
above but including spectroscopically--confirmed galaxies within the
interval and excluding galaxies with spectroscopic redshifts outside
the interval.  The result is a similar but not identical density
map. Noteworthy is the more pronounced concentration of photometric
members, which are defined to be members according to photometric
redshifts, in the western part of the field, which is close to the
three massive, red $z=1.6$ galaxies confirmed by the K20 survey
\citep[][ just to the west of the GMASS field]{cim04}. The circle on
this figure indicates the extent of the high--density region defined
in Sect.~\ref{ssec:peak_intro}. The radius of this circle is chosen
to be 1 (physical) Mpc, similar to the size of low--redshift
clusters. The area of this circular region is similar to the area of a
square defined by \citet{cas07} to contain the highest surface density
of $z=1.61$ galaxies. The \emph{high--density region} contains 21 of
the member galaxies; the same number is present in the area outside
this region, which is five times larger and is therefore referred to
as the \emph{low--density region}.

Selection biases in the angular distribution of galaxies with
spectroscopic redshifts, can be caused by bright objects preventing
the detection of faint objects, for example distant galaxies, within a
certain radial distance. Biases are also introduced by the design of
the spectroscopic masks, which have certain geometrical constraints
such as the smallest possible distance between objects included in
different slits. To estimate the significance of the angular
inhomogeneity displayed by the members with redshifts determined by
GMASS, we carried out a counts-in-cells analysis. In each cell, we
compared the total number of objects targeted in the slits of all six
GMASS masks with the number of $z=1.6$ galaxies observed in that
cell. This analysis was completed for a range of cell sizes, and
showed that the angular distribution of the slits used for
spectroscopy was rather homogeneous. We therefore consider any
observed clustering of members on the sky as true angular clustering.

\subsection{Velocity dispersion}\label{ssec:velocitydispersion}

A fundamental property of galaxy clusters, groups, and their
progenitors is the velocity dispersion of their members. In virialized
systems, this provides a measure of the mass contained within the
system.  A widely adopted estimator of cluster velocity dispersion is the
biweight statistic discussed by \citet[][ hereinafter
BFG90]{bee90}. For a typical dataset of 20 to 50 galaxies, BFG90
showed that the biweight--scale estimator is a robust estimator of
distribution spread and demonstrated its insensitivity to outliers. We
determined the median redshift and biweight statistic of the $z=1.6$
spike in an iterative way. We started the iterative process using all
galaxies in the GMASS field with spectroscopic redshifts in the
interval $1.50 < z < 1.72$. The median $z_{\rm med}$ for these
galaxies was determined, and velocities in the cluster reference frame
were computed as follows:
\begin{equation}
v_{{\rm rest},i} = {\rm c} \times \Delta z \,/\, (1 + z_{\rm med}),
\end{equation}
where c is the speed of light. An estimate of the velocity dispersion,
$\sigma_{\rm spike}$, of the spike was provided by the biweight
statistic for all $v_{{\rm rest},i}$. Subsequently, only galaxies
within the velocity interval $-3\sigma_{\rm spike} \leq v_{{\rm
rest},i} \leq 3\sigma_{\rm spike}$ were included and the computation
of the median redshift and velocity dispersion was repeated until no
changes in the galaxies included occurred. This happened after two
iterations; the 42 galaxies remaining inside the velocity interval at
this stage, were defined to be members of the redshift spike. Their
coordinates, GMASS IDs, and other properties are listed in
Table~\ref{table:spike_gals}.

Our measurements of cluster redshift and velocity dispersion are
$z_{\rm med} = 1.610$ and $\sigma_{\rm spike} = 440^{+95}_{-60}$
km\,s$^{-1}$, respectively. The uncertainty in $\sigma_{\rm spike}$
was taken to be the $1\sigma$ bootstrap confidence interval estimated
from 1000 Monte Carlo simulations\footnote{The computation of
  $\sigma_{\rm spike}$ and its asymmetric error bars was carried out
  employing the ROSTAT software package written and provided by
  BFG90.}. Of the 42 members, 32 have secure redshifts determined by
GMASS and 7 have secure redshifts determined by ESO/GOODS. The
remaining three redshifts are less secure, but removing these objects
from the sample does not alter our final results. Expanding the
redshift interval to a width of $\pm 10\sigma_{\rm spike}$ does not
add any members to the 42 found. In the second inset of
Fig.~\ref{fig:zdistr_dz0.02}, we overlay a Gaussian function that
corresponds to the computed $\sigma_{\rm spike}$. The skewness and
kurtosis of the redshifts distribution are -0.23 and 3.6,
respectively. As is evident in the histogram, there is an enhancement
of values, relative to a Gaussian, below the distribution mean and the
distribution is much more heavy--tailed than a Gaussian
\citep{bir93}. A Gaussian function may not therefore be the most
suitable description of the redshift distribution. This is also shown
by the more robust estimators of asymmetry and tail index
\citep{bir93}, based on order statistics, which have values of -0.27
and 2.6, respectively, indicating deviation from Gaussianity with
significances of $>$20\% and $>$99\%. Considering only the 21 galaxies
within the high density region (dashed histogram in
Fig.~\ref{fig:zdistr_dz0.02}), we obtain a velocity dispersion of
$\sigma_{\rm high} = 500^{+100}_{-100}$ km\,s$^{-1}$.  As this value
is, within the errors, consistent with the value obtained above for
the 42 members, we use the value of 500\,km\,s$^{-1}$ for the
overdensity's velocity dispersion in the remainder of this paper.

In redshift space, the members seem to form a bimodal
distribution (Fig.~\ref{fig:zdistr_dz0.02}), with a main peak at
$z=1.610$ and a secondary peak at $z=1.602$. A k-means clustering
  analysis indeed shows that, if the presence of two redshift groups
  is assumed, the most likely partitioning is into a small group of
  seven to ten members at a mean redshift $z=1.603$ and a large group
  of the remaining members with mean redshift $z=1.611$.  We also
  fitted two models to the redshift distribution around $z=1.61$, the
  first consisting of a single Gaussian function and the second
  composed of the sum of two Gaussian functions.  In both cases, the
  area under the curve was fixed to be the area given by the histogram
  of the 42 member redshifts at $z=1.6$, resulting in two and five
  free fitting parameters, respectively.  The model comprising the
  summed Gaussian functions, shown in Fig.~\ref{fig:zdistr_dz0.02},
  fits the redshift distribution more accurately: an F-test comparing
  these two models excludes the possibility that the 32\% decrease in
  the value of $\chi^2$ is solely due to chance at the 98.5\% level.

The two peaks might be populated by spatially--separated groups. We
applied a statistical test proposed by \citet{dre88} to quantify
possible spatial substructure using the positions of the galaxy members
and their recessional velocities, but found no significant evidence
that the overdensity is substructured.  \citet{pin96} evaluate
  statistical tests for substructure in clusters of galaxies,
  including the Lee2D \citep{fit88} and Lee3D (\citeauthor{pin96})
  tests.  They find that these tests are less sensitive to
  subclustering than the test introduced by \citeauthor{dre88} but have
  the advantage of being insensitive to non-substructure
  configurations that appear as substructure in other tests (such as
  elongation and a velocity dispersion gradient).  We applied
  both tests to our data and found marginal evidence of substructure using
  the Lee2D test and no evidence using the Lee3D test.  The tests
  infer a consistent angle for the line, on which the projected
  member positions show the largest contrast at $\sim75^\circ$ w.r.t.\
  the line of constant declination. As this angle is almost identical
  to the position angle of the GMASS field plotted on
  Fig.~\ref{fig:geom}, we do not indicate this angle separately.  We
  note that the six masks used for the GMASS spectroscopy all had
  different position angles and the similarity between the two angles
  described above is considered to be a coincidence.  Finally, the
two-dimensional two-sided K-S test \citep{fas87} also showed that the
angular distribution of the eleven $1.600 < z \leq 1.608$ galaxies
does not differ significantly from that of the 31 galaxies in the
interval $1.608 < z < 1.622$.  The fact that no clear separation
  between the two redshift groups is seen in projection does not
  exclude the existence of substructure along the line of sight.

For a relaxed cluster, the virial radius is similar to $R_{200}$, the
radius inside which the density is 200 times the critical density.
Using the redshift dependence of the critical density and the virial
mass to relate the line-of-sight velocity dispersion to the cluster
mass, $R_{200}$ can be expressed as \citep[see][]{fin05}
\begin{equation}
R_{200} = 2.47\, {\sigma \over 1000\, {\rm km\, s^{-1}}}\, 
{1 \over \sqrt{\Omega_\Lambda + \Omega_{\rm m} (1 + z)^3}} \,
h_{70}^{-1}\,{\rm Mpc.}
\end{equation}
In addition, the virial mass of a cluster can be expressed as a
function of velocity dispersion using the above equation.  Assuming that
the virial radius is equal to $R_{200}$, one obtains \citep[see also][]{fin05}
\begin{equation}
M_{\rm vir} = {3 \, \sigma \, R_{200} \over G },
\end{equation}
and
\begin{equation}
M_{\rm vir} = 1.7\times10^{15}\, 
{\left( \sigma \over 1000\, {\rm km\, s^{-1}}\right)^3}\, 
{1 \over \sqrt{\Omega_\Lambda + \Omega_{\rm m} (1 + z)^3}} \,
h_{70}^{-1}\,M_\odot.
\end{equation}
Although we do not know whether the structure under study here is
virialized, we do apply these equations to derive an indication of the
virial radius and mass, obtaining (for $\sigma$ = 500 km s$^{-1}$)
$R_{200}$ = 0.5 Mpc and $M_{\rm vir}$ = 9 $\times 10^{13}$ M$_\odot$.

\subsection{Galaxy overdensity and mass contained}

The galaxy overdensity in redshift space is defined to be the number
of galaxies found in a certain redshift interval, divided by the
number of galaxies expected in this interval minus one. To determine a
galaxy overdensity in redshift space, one has to assume an expected
number of galaxies in this interval. It is obviously impossible to
estimate the expected number of galaxies without introducing any
biases such as those discussed in
Sect.~\ref{sec:spectroscopic_bias}. In the absence of a superior
method, we assume a flat n$(z)$ distribution and employ the
108 galaxies with spectroscopically--confirmed redshifts within the
GMASS sample in the range $z_1 = 1.400 < z < 1.900 = z_2$.  The lower
limit is set by the selection criteria for GMASS spectroscopic
observations and the upper limit is based on the appearance of the
redshift distribution shown in Fig.~\ref{fig:zdistr_dz0.02}, which
declines at $z > 2$. Indeed the number of galaxies expected varies by
less than 5\% for $1.75 < z_2 < 2.00$.  
For the redshift interval $1.601 < z < 1.621$ with $\Delta z =
  0.02$, 4.32 galaxies would be expected but 40 are observed
  (subtracting the two galaxies that define the boundaries of this
  interval).  Therefore, we derive an estimated overdensity of
  $40/4.32-1=8.3$$\pm$1.5, where the uncertainty is based on Poisson
  statistics.  We note that restricting the redshift interval of the
  peak to $1.601 < z < 1.615$, i.e., excluding a single galaxy, leads
  to an estimated overdensity of $39/3.02-1=11.9$$\pm$2.2.  In
  addition, excluding the redshift interval $1.572 < z < 1.670$ from
  the n$(z)$ count, which includes all member galaxies and additional
  empty redshift space, the estimated overdensity would be
  $40/3.28-1=11.2$$\pm$2.2.  In the following, we conservatively adopt
  an overdensity of 8.3$\pm$1.5.

To estimate the total mass contained within the volume occupied by the
spike, we need to measure this volume and the matter overdensity,
which is related to the galaxy overdensity by the bias factor
\begin{equation}
1 + b\,\delta_{\rm mass} = C (1 + \delta_{\rm gal,obs})
\end{equation}
\citep[e.g.,][]{ste98}. The bias factor is plausibly 1.6, as
  extrapolated from measurements up to $z = 1.5$ by \citet{mar05b},
  which is lower than the range of 2--4 assumed by \citet{ste98}, who
  originally introduced the method followed below.  Assuming this bias
  factor, the matter overdensity then equals $\delta_{\rm mass} =
  2.45^{+0.25}_{-0.30}$. Here, we assume the observed galaxy
  overdensity $\delta_{\rm gal,obs}=8.3\pm1.5$, which implies a value
  of $0.53^{-0.04}_{-0.04}$ for the factor C and takes into
account the redshift-space distortions caused by the peculiar
velocities of the galaxies \citep{ste98}. In the transverse direction,
it is difficult to estimate the spatial extent of the structure
because the GMASS field does not cover the entire overdensity.  We
adopt two approaches here: in the first approach, we follow
\citet{ste98} and similar work \citep{ven07,ota08}, by taking the
entire field sampled, while in the second, more conservative and
  maybe more accurate, approach, we take the localized high density
region only as defined by a 1 Mpc (physical) radius from the central
galaxy. In the first case, from the observed redshift depth of the
overdensity of $\delta z = 0.016$, excluding the single galaxy at
$z=1.621$, and taking into account that $V_{\rm true} = V_{\rm obs} /
C$, the mass enclosed in this volume is $\bar{\rho}_{\rm m} V (1
  + \delta_{\rm m})= 5.9\pm0.9 \times 10^{14} $M$_\odot$. The volume
considered here, which is equivalent to a sphere of a $\sim$10\,Mpc
comoving radius, is large, corresponding to a physical size of
$\sim$3.8\,Mpc at $z=1.61$. In the second case, the physical volume of
a sphere has a physical radius of 1\,Mpc, but the overdensity is
almost six times larger. The mass contained in this smaller
  volume is therefore a fraction 0.018$\times$6 of that stated above,
  or $6.4\pm1.0 \times 10^{13} $M$_\odot$.

The total stellar mass contained within the 21 galaxies inside the
high--density region equals $\sim 6 \times 10^{11}$ M$_\odot$ (the 21
galaxies inside the low density region add only another $\sim 2 \times
10^{11}$ M$_\odot$). Applying a total-to-stellar-mass ratio of 100,
appropriate for clusters with a velocity dispersion in the range
425--800 km\,s$^{-1}$ \citep{bal07}, in a similar way to the approach
of \citet{mcc07} for a structure at $z=1.5$, we obtain a total mass of
$\sim 6 \times 10^{13}$ M$_\odot$, which is consistent with the mass obtained
above for the sphere circumscribing the high--density region.

\subsection{X-ray emission}

\citet{cas07} reported three X-ray sources within the high--density
region, one coinciding with the peak of their galaxy density contours
and including a late--type galaxy of a peculiar morphology. We
identify this galaxy with GMASS\,2540, which indeed has a redshift $z
= 1.613$, measured using GMASS spectroscopy. In addition, several
other galaxies with spectroscopic redshifts at $z = 1.6$ from GMASS
are located within the soft X-ray contours displayed in Fig.\ 2 of
\citeauthor{cas07}, namely GMASS\,2180, 2286, and 2251 at $z = 1.608$,
1.604, and 1.609, respectively.

The CDFS was observed by \emph{Chandra} for 1\,Msec \citep{ros02},
which allows detection of extended emission down to $1\times 10^{-16}$
erg\,cm$^{-2}$\,s$^{-1}$ in the 0.5--2 keV band for relatively compact
sources.  More recently, these observations were extended to 2\,Msec
\citep{luo08}.  In the GMASS region, the Chandra PSF FWHM is 2--3
arcsec and the median effective exposure time at the positions of the
overdensity members is 1.56\,Msec. For sources with an extent of 5--10
arcsec, which are easily resolved by Chandra, the 2\,Msec data have a
detection limit (S/N $\approx 3$) of $6\times 10^{-17}$
erg\,cm$^{-2}$\,s$^{-1}$ in the soft band.  Given the size of the
GMASS field and the cumulative X--ray cluster number counts presented
in \citet{ros02b}, at least $\sim\! 4$ low X--ray luminosity clusters
(at any redshift) are expected to be detected. The presence of a
galaxy group or cluster in this field is therefore quite likely.

At $z=1.6$, this X-ray sensitivity implies a limit to the X-ray
luminosity of $1\times 10^{42}$ erg s$^{-1}$ in the observed frame and
0.5--2 keV band. For a T$=2$ keV group, the K-correction amounts to 9
and the bolometric correction to 2.3. Therefore, the Chandra
observations imply a limit of $L_{\rm X, bol}=2.1\times 10^{43}$ erg
s$^{-1}$. \citet{mah01} provide an empirical relation $L_X=5\times
10^{42}$ erg s$^{-1}$ $(\sigma/330)^{4.4}$, which has significant
scatter, in particular for groups. For the measured velocity
dispersion of the redshift spike of 500$\pm$100 km s$^{-1}$, one would
expect 
an X-ray luminosity $3_{-2}^{+3} \times 10^{43}$ erg s$^{-1}$; this is
consistent with our limit because of both the large scatter in the
$L_X$ -- $\sigma$ relation, and the error in the velocity dispersion,
which may have been overestimated if the overdensity is not yet
virialized. An extended X-ray source is detected neither in the
vicinity of any ETG in the north of the field, nor around the triplet
clump. In addition, by stacking X-ray photons in the 0.5--2 keV band
around the seven ETGs, for a total effective exposure time of
11.3\,Msec, no signal is detected down to $0.7\times 10^{-17}$ erg
cm$^{-2}$ s$^{-1}$.

\subsection{Colour--magnitude diagram}\label{ssec:cm}

Although more pronounced in galaxy clusters, the field population of
galaxies up to at least $z \sim 1$ shows a bimodal distribution in
colour, with red galaxies on the passive side and blue galaxies on the
active side, separated by a green valley of galaxies in transition
from blue to red \citep[e.g.,][]{bel04b}.  The emergence of colour
bimodality at $z \sim 2$ is the subject of another paper employing the
GMASS spectra \citep{cas08}.  Within cluster populations, the
bimodality is detected as a 
concentration of red (early--type) galaxies on a line in a
colour-magnitude diagram using colours that straddle the 4000\,\AA\
break, which is referred to as the red sequence (RS).  This sequence
is more pronounced in regions of high galaxy density (i.e., clusters
and groups) because galaxies appear to evolve more rapidly in such an
environment; \citet{tho05}, for example, found that most of the
star--formation activity in early--type galaxies probably occurred at
$3 < z < 5$ in high, and at $1 < z < 2$ in low density environments.
The RS was found to be present in a cluster at $z = 0.83$
\citep{tra07}, has been used to find clusters up to $z \sim 1.4$
\citep{gla00}, and there is evidence that the bright end (M$_* >
10^{11}$\,M$_\odot$) of the RS is populated up to $z \sim 2$ in
massive cluster progenitors \citep{kod07}.

\begin{figure}
  \centering
  \includegraphics[width=\columnwidth]{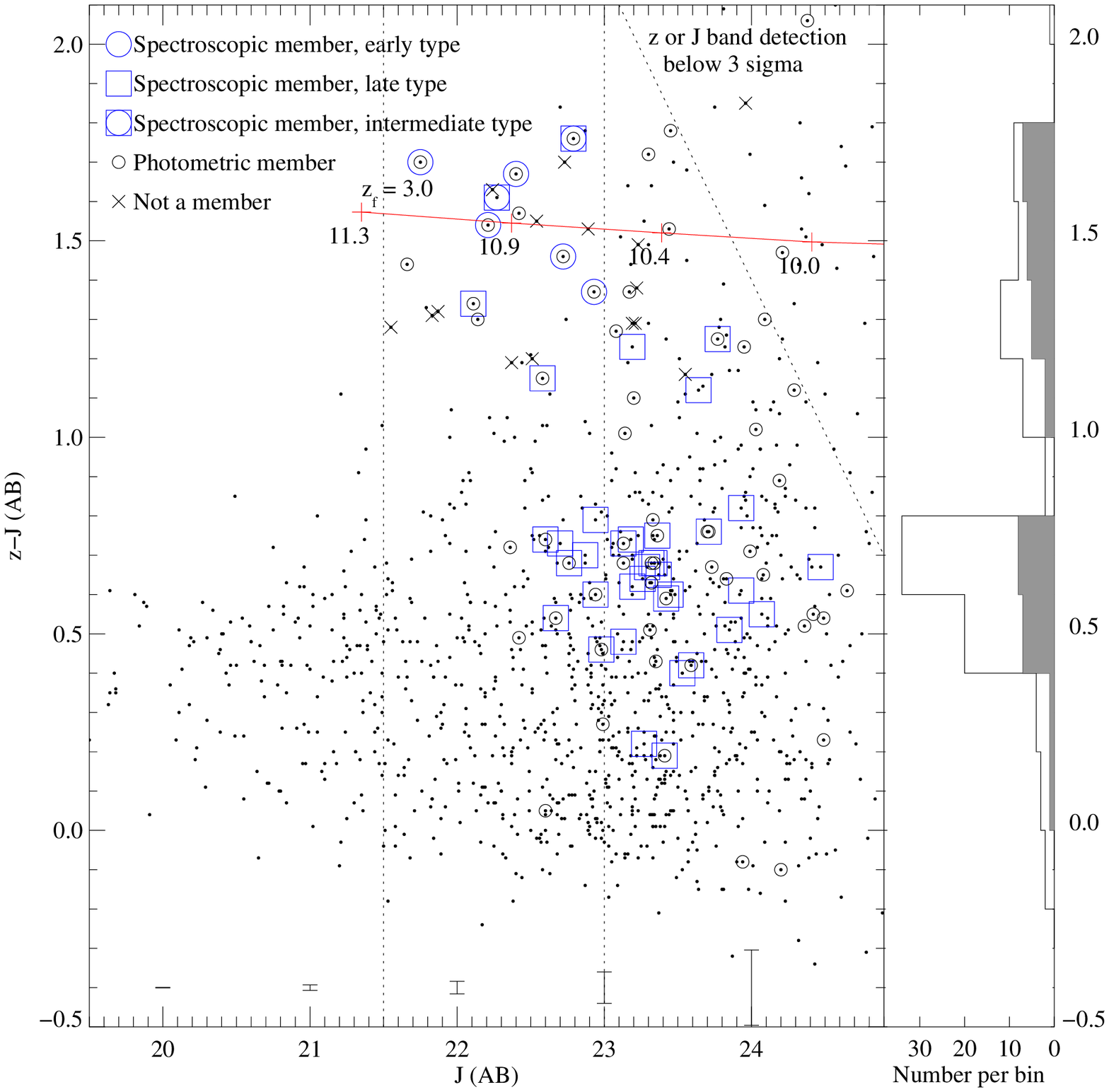}

  \caption{Colour-magnitude diagram of all objects in the GMASS
    catalogue (excluding those undetected in one of the two bands)
    showing $z-J$ colour vs. $J$ magnitude.  Large symbols indicate
    galaxies at $1.600 < z_{\rm spec} < 1.622$, small circles indicate
    galaxies at $1.50 < z_{\rm phot} < 1.70$ (not including those with
    $z_{\rm spec} < 1.600$ or $z_{\rm spec} > 1.622$).  Galaxies with
    spectroscopic redshifts outside the range $1.600 < z_{\rm spec} <
    1.622$ are indicated by crosses. Spectroscopic early types are
    indicated by circles (5), intermediate types by a both circles and
    boxes (2), and late types by boxes only (35).  The solid (red)
    line indicates a theoretical red sequence observed at $z=1.60$ for
    elliptical galaxies with a range of masses (as indicated in
    logarithmic solar masses) formed at $z_{\rm f} = 3.0$.  At the
    bottom of the plot, typical errors in colour are indicated for
    each $J$ magnitude.  In addition, the objects to the right of the
    diagonal dashed line have data of signal-to-noise ratios below
    three.  To the right, a histogram is displayed for all member
    galaxies (open histogram) and for members within the range $21.5 <
    J < 23.0$ (indicated by dashed vertical lines in the
    colour-magnitude diagram) where the confirmed red sequence
    galaxies are located (filled histogram).}
  \label{fig:cm_diagram}

\end{figure}

For $z = 1.6$, the 4000\,\AA\ break is shifted to 10\,400\,\AA, between
the $z$ and $J$ bands.  We plot a $z-J$ versus $J$ diagram in
Fig.~\ref{fig:cm_diagram}, indicating the member galaxies by large
symbols (defined according to spectral class: circles for early types,
diamonds for late types, and both for intermediate types).  Also
indicated, with small circles, are objects without a spectroscopic
redshift but with a photometric redshift in the range $1.50 < z_{\rm
  phot} < 1.70$.  There appear to be two populations of spectroscopic
members: a blue population around $z-J = 0.6$, and a red population
with $z-J \sim 1.5$.  Clearly the early and intermediate types
(derived from the spectra) have the reddest $z-J$ colours: all are at
$z-J > 1.3$.  The figure shows that, although there are many member
candidates present at $z-J > 1.3$, there is also some contamination
from objects (as indicated by the crosses) with spectroscopic
redshifts outside the range $1.600 < z_{\rm spec} < 1.622$.

\section{Photometric, morphological and spectral, galaxy properties in
  spike and field}\label{sec:galaxyproperties}

In this section, we compare the galaxy properties of samples of member
and field galaxies.  The properties include the galaxy morphology,
spectral type, fitted SED, and several properties derived from this
SED.  We refer to \citet{cas08} and \citet{cim08} for more details
about the morphological analysis and SED fitting.

\begin{table*}
\caption{Spike galaxy redshifts, photometry, properties derived from fitted SEDs, and morphological classifications}
\label{table:spike_gals}
\centering
\begin{tabular}{rrlrrrrrrrrrr}
\hline\hline
\multicolumn{1}{c}{ID} & \multicolumn{1}{c}{R.A. \& Dec.} &  \multicolumn{1}{c}{$z^{\mathrm{a}}$} & O$^{\mathrm{b}}$ & Cl$^{\mathrm{c}}$ & 4.5$\mu$m & $z'$ & $J$ & $M_*^{\mathrm{d}}$ & SFR$^{\mathrm{e}}$ & Age$^{\mathrm{f}}$ & A$_{\rm V}$ & M$^{\mathrm{g}}$\\
\hline
 365 & 03 32 27.80  -27 48 12.0 &1.609 & 1 &2.0 & 23.10$\pm$0.02 & 24.4$\pm$0.1 & 23.9$\pm$0.1 &  9.3$^{+0.1}_{-0.2}$& 10$^{+06}_{-03}$& 0.26$^{+0.25}_{-0.16}$&  0.3$^{+0.2}_{-0.2}$& 2 \\
 648 & 03 32 29.81  -27 47 19.7 &1.609 & 2 &2.0 & 22.20$\pm$0.01 & 23.9$\pm$0.1 & 23.2$\pm$0.1 &  9.7$^{+0.1}_{-0.1}$& 25$^{+07}_{-06}$& 0.13$^{+0.03}_{-0.03}$&  0.9$^{+0.1}_{-0.1}$& 2 \\
 781 & 03 32 17.71  -27 47 02.9 &1.605 & 3 &2.0 & 22.54$\pm$0.02 & 23.5$\pm$0.1 & 23.3$\pm$0.1 &  9.5$^{+0.1}_{-0.1}$&  8$^{+01}_{-01}$& 0.57$^{+0.01}_{-0.06}$&  0.0$^{+0.1}_{-0.0}$& 3 \\
1254 & 03 32 20.17  -27 45 49.3 &1.610 & 1 &2.0 & 22.21$\pm$0.02 & 23.6$\pm$0.1 & 23.4$\pm$0.1 &  9.4$^{+0.1}_{-0.1}$& 34$^{+01}_{-01}$& 0.09$^{+0.01}_{-0.01}$&  0.6$^{+0.1}_{-0.1}$& 1 \\
1282 & 03 32 33.86  -27 45 42.6 &1.621 & 2 &2.0 & 21.50$\pm$0.01 & 23.9$\pm$0.1 & 23.1$\pm$0.1 &  9.9$^{+0.1}_{-0.1}$& 51$^{+20}_{-10}$& 0.11$^{+0.05}_{-0.02}$&  1.5$^{+0.1}_{-0.1}$& 2 \\
1295 & 03 32 28.59  -27 45 42.2 &1.611 & 2 &2.0 & 23.08$\pm$0.03 & 24.5$\pm$0.1 & 23.7$\pm$0.1 &  9.5$^{+0.1}_{-0.1}$&  4$^{+03}_{-01}$& 0.23$^{+0.03}_{-0.07}$&  0.3$^{+0.2}_{-0.1}$& 3 \\
1380 & 03 32 25.25  -27 45 29.0 &1.612 & 3 &2.0 & 22.30$\pm$0.02 & 24.1$\pm$0.1 & 23.4$\pm$0.1 &  9.8$^{+0.1}_{-0.2}$& 16$^{+11}_{-02}$& 0.29$^{+0.03}_{-0.17}$&  0.8$^{+0.2}_{-0.1}$& 3 \\
1495 & 03 32 35.36  -27 45 12.6 &1.611 & 1 &2.0 & 21.20$\pm$0.01 & 24.0$\pm$0.1 & 23.3$\pm$0.1 &  9.9$^{+0.1}_{-0.1}$& 97$^{+03}_{-05}$& 0.10$^{+0.01}_{-0.01}$&  1.6$^{+0.1}_{-0.1}$& 3 \\
1556 & 03 32 37.74  -27 45 05.5 &1.607: & 2 &2.0 & 21.03$\pm$0.01 & 24.8$\pm$0.2 & 23.6$\pm$0.1 & 10.4$^{+0.1}_{-0.1}$& 49$^{+11}_{-27}$& 0.20$^{+0.08}_{-0.02}$&  2.2$^{+0.1}_{-0.3}$& 3 \\
1667 & 03 32 40.99  -27 44 50.2 &1.613 & 1 &2.0 & 22.51$\pm$0.02 & 25.1$\pm$0.2 & 24.5$\pm$0.1 &  9.4$^{+0.1}_{-0.1}$& 33$^{+01}_{-05}$& 0.09$^{+0.02}_{-0.01}$&  1.8$^{+0.1}_{-0.1}$& 3 \\
1691 & 03 32 31.90  -27 44 45.0 &1.612 & 1 &2.0 & 21.11$\pm$0.01 & 23.6$\pm$0.1 & 22.9$\pm$0.1 & 10.1$^{+0.1}_{-0.1}$& 84$^{+04}_{-19}$& 0.10$^{+0.03}_{-0.01}$&  1.5$^{+0.1}_{-0.1}$& 3 \\
1708 & 03 32 23.12  -27 44 42.2 &1.607 & 1 &2.0 & 20.81$\pm$0.01 & 25.0$\pm$0.2 & 23.8$\pm$0.1 & 10.5$^{+0.1}_{-0.1}$& 73$^{+26}_{-34}$& 0.18$^{+0.07}_{-0.04}$&  2.5$^{+0.1}_{-0.2}$& 3 \\
1808 & 03 32 26.15  -27 44 33.3 &1.609 & 1 &2.0 & 22.61$\pm$0.02 & 23.9$\pm$0.1 & 23.5$\pm$0.1 &  9.4$^{+0.1}_{-0.1}$& 29$^{+01}_{-13}$& 0.10$^{+0.08}_{-0.01}$&  1.0$^{+0.1}_{-0.2}$& 2 \\
1871 & 03 32 26.82  -27 44 24.7 &1.613 & 2 &2.0 & 22.66$\pm$0.02 & 23.4$\pm$0.1 & 23.0$\pm$0.1 &  9.5$^{+0.1}_{-0.1}$&  9$^{+01}_{-01}$& 0.40$^{+0.01}_{-0.04}$&  0.0$^{+0.1}_{-0.0}$& 3 \\
1979 & 03 32 24.64  -27 44 07.8 &1.611 & 1 &2.0 & 20.95$\pm$0.01 & 23.2$\pm$0.1 & 22.7$\pm$0.1 & 10.0$^{+0.1}_{-0.1}$&116$^{+04}_{-03}$& 0.09$^{+0.02}_{-0.01}$&  1.2$^{+0.1}_{-0.1}$& 3 \\
1988 & 03 32 24.02  -27 44 08.3 &1.610 & 2 &2.0 & 22.77$\pm$0.03 & 24.0$\pm$0.1 & 23.6$\pm$0.1 &  9.4$^{+0.1}_{-0.2}$& 11$^{+07}_{-02}$& 0.29$^{+0.12}_{-0.20}$&  0.4$^{+0.2}_{-0.1}$& 2 \\
2055 & 03 32 26.77  -27 43 58.1 &1.611 & 1 &2.0 & 22.21$\pm$0.02 & 24.8$\pm$0.2 & 23.9$\pm$0.1 &  9.8$^{+0.1}_{-0.1}$& 20$^{+12}_{-07}$& 0.26$^{+0.11}_{-0.14}$&  1.3$^{+0.2}_{-0.2}$& 2 \\
2081 & 03 32 29.86  -27 43 54.8 &1.601 & 1 &2.0 & 22.09$\pm$0.01 & 24.1$\pm$0.1 & 23.4$\pm$0.1 &  9.6$^{+0.1}_{-0.1}$&  5$^{+01}_{-01}$& 1.02$^{+0.12}_{-0.11}$&  0.1$^{+0.1}_{-0.1}$& 2 \\
2111 & 03 32 27.94  -27 42 45.7 &1.610 & 1 &1.0 & 20.64$\pm$0.01 & 24.1$\pm$0.1 & 22.4$\pm$0.1 & 10.6$^{+0.1}_{-0.1}$&  0$^{+01}_{-00}$& 1.14$^{+0.06}_{-0.06}$&  0.0$^{+0.1}_{-0.0}$& 1 \\
2113 & 03 32 22.00  -27 42 43.5 &1.613 & 1 &2.0 & 20.37$\pm$0.01 & 24.4$\pm$0.1 & 23.2$\pm$0.1 & 10.6$^{+0.1}_{-0.1}$& 84$^{+21}_{-45}$& 0.45$^{+0.56}_{-0.05}$&  1.9$^{+0.1}_{-0.4}$& 3 \\
2142 & 03 32 23.54  -27 42 49.3 &1.610 & 1 &2.0 & 22.22$\pm$0.02 & 24.0$\pm$0.1 & 23.3$\pm$0.1 &  9.8$^{+0.1}_{-0.1}$& 10$^{+02}_{-04}$& 0.23$^{+0.06}_{-0.02}$&  0.7$^{+0.1}_{-0.2}$& 2 \\
2148 & 03 32 36.30  -27 42 49.5 &1.609 & 1 &1.0 & 19.80$\pm$0.01 & 23.4$\pm$0.1 & 21.8$\pm$0.1 & 11.0$^{+0.1}_{-0.1}$&  0$^{+01}_{-00}$& 1.28$^{+0.01}_{-0.14}$&  0.1$^{+0.1}_{-0.1}$& 1 \\
2180 & 03 32 29.56  -27 42 56.0 &1.608 & 1 &2.0 & 21.00$\pm$0.01 & 23.4$\pm$0.1 & 22.7$\pm$0.1 & 10.2$^{+0.1}_{-0.1}$& 75$^{+21}_{-09}$& 0.13$^{+0.01}_{-0.03}$&  1.4$^{+0.1}_{-0.1}$& 2 \\
2196 & 03 32 36.67  -27 42 58.5 &1.614 & 1 &1.0 & 20.34$\pm$0.01 & 23.8$\pm$0.1 & 22.2$\pm$0.1 & 10.7$^{+0.1}_{-0.1}$&  0$^{+01}_{-00}$& 1.14$^{+0.14}_{-0.01}$&  0.0$^{+0.1}_{-0.0}$& 1 \\
2251 & 03 32 29.48  -27 43 22.0 &1.609 & 1 &2.0 & 20.64$\pm$0.01 & 23.7$\pm$0.1 & 22.6$\pm$0.1 & 10.5$^{+0.1}_{-0.1}$&  7$^{+06}_{-01}$& 0.90$^{+0.04}_{-0.15}$&  0.5$^{+0.2}_{-0.1}$& 2 \\
2286 & 03 32 29.99  -27 43 22.6 &1.604 & 1 &1.0 & 20.82$\pm$0.01 & 24.3$\pm$0.1 & 22.9$\pm$0.1 & 10.5$^{+0.1}_{-0.1}$&  0$^{+01}_{-00}$& 1.02$^{+0.12}_{-0.08}$&  0.2$^{+0.1}_{-0.1}$& 1 \\
2341 & 03 32 17.52  -27 43 36.6 &1.602 & 3 &2.0 & 22.47$\pm$0.02 & 24.0$\pm$0.1 & 23.4$\pm$0.1 &  9.6$^{+0.1}_{-0.1}$&  7$^{+02}_{-01}$& 0.81$^{+0.33}_{-0.35}$&  0.2$^{+0.1}_{-0.1}$& 2 \\
2352 & 03 32 33.88  -27 42 04.1 &1.604 & 1 &2.0 & 20.09$\pm$0.01 & 23.4$\pm$0.1 & 22.1$\pm$0.1 & 10.8$^{+0.1}_{-0.1}$& 22$^{+08}_{-14}$& 0.36$^{+0.09}_{-0.03}$&  1.0$^{+0.1}_{-0.4}$& 2 \\
2355 & 03 32 14.32  -27 43 32.9 &1.610 & 1 &1.0 & 20.92$\pm$0.01 & 24.2$\pm$0.1 & 22.7$\pm$0.1 & 10.4$^{+0.1}_{-0.1}$&  0$^{+01}_{-00}$& 0.90$^{+0.01}_{-0.10}$&  0.0$^{+0.1}_{-0.0}$& 1 \\
2361 & 03 32 26.05  -27 42 36.6 &1.609 & 3 &1.5 & 20.40$\pm$0.01 & 23.9$\pm$0.1 & 22.3$\pm$0.1 & 10.8$^{+0.1}_{-0.1}$&  0$^{+01}_{-00}$& 1.43$^{+0.27}_{-0.01}$&  0.2$^{+0.1}_{-0.1}$& 1 \\
2368 & 03 32 17.10  -27 43 41.9 &1.612 & 1 &2.0 & 21.15$\pm$0.01 & 24.0$\pm$0.1 & 23.3$\pm$0.1 & 10.0$^{+0.3}_{-0.1}$&108$^{+03}_{-65}$& 0.10$^{+0.41}_{-0.01}$&  1.6$^{+0.1}_{-0.4}$& 2 \\
2438 & 03 32 26.41  -27 42 28.4 &1.615 & 2 &2.0 & 20.87$\pm$0.01 & 23.5$\pm$0.1 & 22.9$\pm$0.1 & 10.2$^{+0.1}_{-0.1}$& 83$^{+25}_{-17}$& 0.26$^{+0.11}_{-0.15}$&  1.4$^{+0.1}_{-0.1}$& 2 \\
2454 & 03 32 28.91  -27 43 03.6 &1.602 & 1 &2.0 & 20.80$\pm$0.01 & 24.0$\pm$0.1 & 23.4$\pm$0.1 & 10.1$^{+0.1}_{-0.1}$&153$^{+02}_{-04}$& 0.09$^{+0.01}_{-0.01}$&  1.8$^{+0.1}_{-0.1}$& 3 \\
2493 & 03 32 38.51  -27 42 28.0 &1.607 & 1 &2.0 & 20.88$\pm$0.01 & 23.3$\pm$0.1 & 22.6$\pm$0.1 & 10.2$^{+0.1}_{-0.1}$&114$^{+03}_{-25}$& 0.10$^{+0.03}_{-0.01}$&  1.4$^{+0.1}_{-0.1}$& 3 \\
2540 & 03 32 30.33  -27 42 40.3 &1.613 & 1 &2.0 & 21.40$\pm$0.01 & 23.7$\pm$0.1 & 22.9$\pm$0.1 & 10.1$^{+0.1}_{-0.1}$& 51$^{+15}_{-10}$& 0.14$^{+0.03}_{-0.03}$&  1.0$^{+0.1}_{-0.1}$& 3 \\
2543 & 03 32 35.92  -27 42 41.0 &1.612 & 1 &1.5 & 20.26$\pm$0.01 & 24.6$\pm$0.1 & 22.8$\pm$0.1 & 10.7$^{+0.1}_{-0.1}$&  0$^{+01}_{-00}$& 1.02$^{+0.09}_{-0.03}$&  0.6$^{+0.1}_{-0.1}$& 1 \\
2550 & 03 32 30.08  -27 42 12.2 &1.601 & 1 &2.0 & 22.03$\pm$0.01 & 23.9$\pm$0.1 & 23.3$\pm$0.1 &  9.7$^{+0.1}_{-0.1}$& 39$^{+04}_{-08}$& 0.09$^{+0.05}_{-0.01}$&  1.0$^{+0.1}_{-0.1}$& 2 \\
2603 & 03 32 27.85  -27 43 05.7 &1.612 & 1 &2.0 & 21.74$\pm$0.01 & 23.6$\pm$0.1 & 23.1$\pm$0.1 &  9.9$^{+0.1}_{-0.1}$& 41$^{+14}_{-05}$& 0.13$^{+0.01}_{-0.04}$&  0.9$^{+0.1}_{-0.1}$& 2 \\
 505 & 03 32 43.37  -27 47 43.5 &1.610: & 2 &2.0 & 22.86$\pm$0.02 & 24.5$\pm$0.1 & 23.9$\pm$0.1 &  9.3$^{+0.2}_{-0.1}$& 15$^{+06}_{-07}$& 0.10$^{+0.13}_{-0.01}$&  0.8$^{+0.1}_{-0.2}$& 3 \\
1081 & 03 32 17.51  -27 46 14.0 &1.610 & 2 &2.0 & 21.03$\pm$0.01 & 23.4$\pm$0.1 & 22.8$\pm$0.1 & 10.0$^{+0.1}_{-0.1}$&125$^{+01}_{-22}$& 0.09$^{+0.02}_{-0.01}$&  1.5$^{+0.1}_{-0.1}$& 3 \\
1399 & 03 32 41.66  -27 45 25.6 &1.615 & 1 &2.0 & 22.71$\pm$0.02 & 24.6$\pm$0.1 & 24.1$\pm$0.1 &  9.3$^{+0.2}_{-0.1}$& 17$^{+07}_{-04}$& 0.09$^{+0.16}_{-0.01}$&  1.0$^{+0.1}_{-0.1}$& 3 \\
1413 & 03 32 24.66  -27 45 21.6 &1.614: & 2 &2.0 & 21.50$\pm$0.01 & 23.8$\pm$0.1 & 23.2$\pm$0.1 &  9.8$^{+0.1}_{-0.1}$& 87$^{+01}_{-01}$& 0.09$^{+0.01}_{-0.01}$&  1.2$^{+0.1}_{-0.1}$& 3 \\
\hline
\end{tabular}
\begin{list}{}{}
\item[$^{\mathrm{a}}$] Spectroscopic redshift, colon indicates uncertain redshift
\item[$^{\mathrm{b}}$] Origin of redshift: (1) GMASS (2) ESO/GOODS
\item[$^{\mathrm{c}}$] Spectral class: (1.0) early type, (1.5) early type with signs of star formation, (2.0) late type
\item[$^{\mathrm{d}}$] Logarithm of stellar mass in solar masses
\item[$^{\mathrm{e}}$] SFR in solar mass per year
\item[$^{\mathrm{f}}$] Age in Gyr
\item[$^{\mathrm{g}}$] Morphology: (1) Elliptical (2) Spiral (3) Irregular
\end{list}
\end{table*}

\begin{figure*} \centering 
  \includegraphics[width=\linewidth]{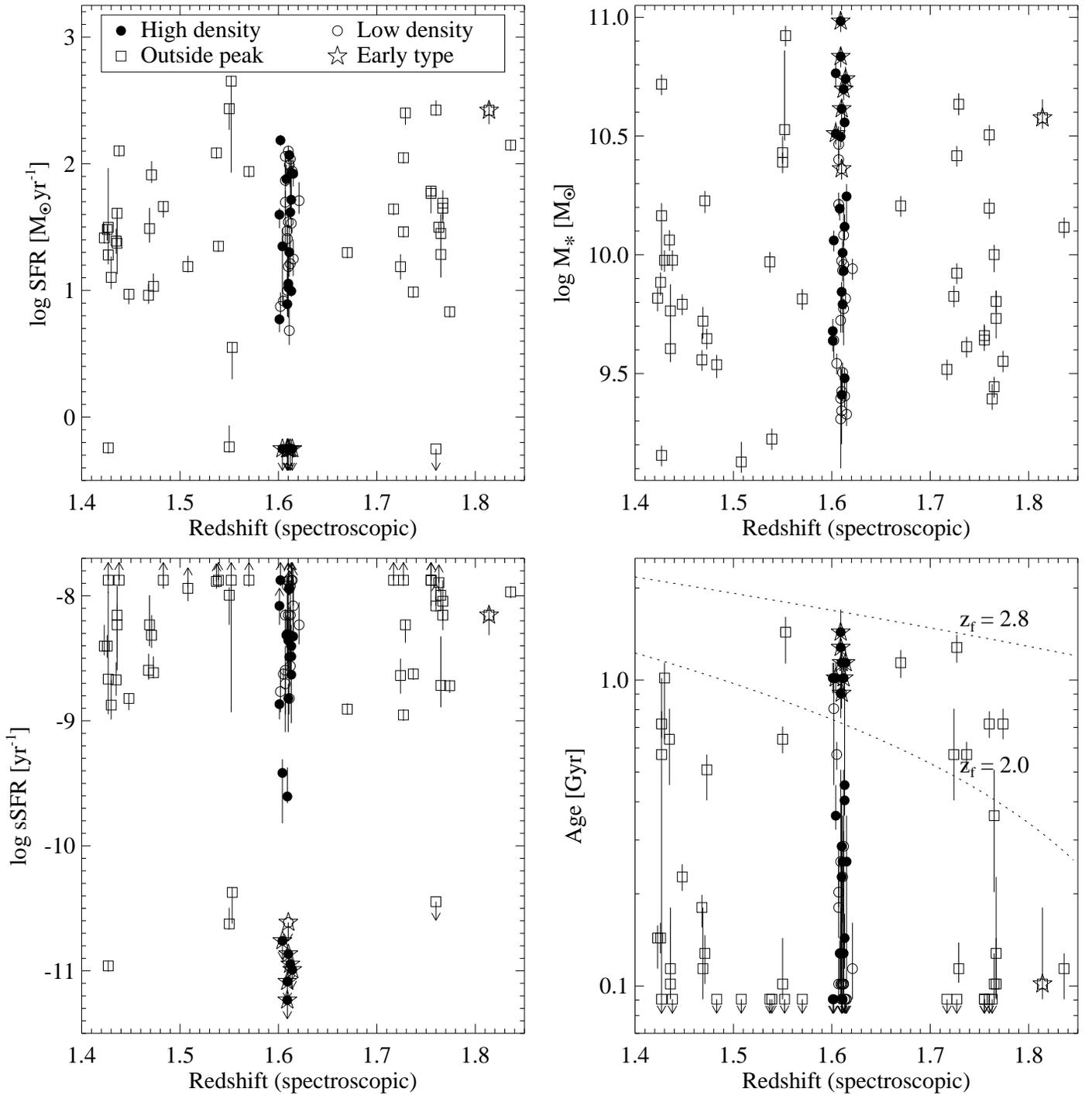}
  \caption{Comparison of galaxy properties for the galaxies in the
    redshift range $1.4 < z < 1.8$.  From left to right and top to
    bottom: SFR, mass, sSFR, and age.  Symbols are for galaxies
    within the high density region (filled circles), low density region
    (open circles), and outside the peak (boxes). Spectroscopic early--type 
    galaxies are indicated by stars. The dashed lines in the 
    fourth panel indicates the age of a galaxy formed with a short burst 
    of star formation at $z = 2.0$ and at $z = 2.8$.
    \label{fig:gal_props_1}}
\end{figure*}

\subsection{Morphological analysis}

The morphological analysis was performed independently by two of
  us (PC and GR) in the ACS F850LP--band image \citep{cas08}. Three
  classes were recognized by eye: (1) elliptical, (2) spiral, and (3)
  irregular.  For the faint galaxies (close to $z=1.6$), the two
  classifiers agreed in more than 70\% of the cases.  The
  discrepancies mainly occur for galaxies close to the boundary
  between spirals and irregulars, whereas agreement was obtained for
  90\% of the early-types.  The final catalogue has been produced by a
  reconciliation process between the two classifiers for the remaining
  30\% of objects.  We also determined morphological parameters by
  fitting the galaxy light profiles (for details of the fitting
  process, see \citeauthor{cas08}).  We compared the classification by
  eye with the machine--determined parameters, obtaining the following
  results: 90\% of the early-type galaxies (12/13) have a S\'ersic
  index $>$ 2, while 88\% of late-types (63/72) have a S\'ersic index
  $<$ 2.

  The morphological classification was carried out using the ACS
  F850LP--band image, i.e., using the rest-frame UV light of galaxies
  at $z=1.6$.  Therefore, surface brightness dimming and morphological
  k--correction may be affecting the morphological classification.  On
  the one hand, ellipticals are expected to remain symmetrical in the
  UV, but fainter because of the k-correction effect.  On the other
  hand, spirals can appear morphologically later because knots of star
  formation brighten in the UV and disks become fainter because of the
  surface brightness dimming.  The combination of these two effects
  should not affect our visual classification, at least in separating
  early- and late-types.  As the galaxies in our samples are observed
  in almost the same rest-frame UV light (the central rest-frame
  wavelength differs from 394 to 338\,nm for redshifts between $z=1.4$
  and $z=1.8$, respectively, while the filter has a width of 47\,nm in
  the rest-frame), these phenomena do not introduce a differential
  effect on the classification of galaxies at $z=1.4$ to $z=1.8$.

\subsection{SED fits}
 The properties of the stellar populations in each galaxy are
  determined from the SED fitting analysis for the spectroscopic
  redshift determined.  The results of the SED fitting are, therefore,
  far more secure than for one where the redshift is unknown and
  therefore a free parameter in the fitting process, as is the case
  for the determination of photometric redshifts.  We adopt the
  libraries of synthetic spectra by \citet[][ M05]{mar05}, who employs
  the \citet{kro01} initial mass function (IMF) and includes the
  thermally pulsing asymptotic giant branch (TP-AGB) phase of stellar
  evolution. This phase is the dominant source of bolometric and
  near-IR radiation for a simple stellar population in the age range
  from 0.2 to 2 Gyr.

  We adopt exponentially declining star--formation histories, i.e.,
  SFR = $({\cal M}/ \tau) exp(-t/ \tau)$ with $\tau=$ 0.1, 0.3, 1, 2,
  3, 5, 10, 15, and 30 Gyr, and, in addition, the case of a constant
  star--formation rate.  Extinction is treated as a free parameter in
  the optimization, by adopting the extinction curve of Calzetti et
  al. (2000).  We assumed solar metallicity for all the models.  The
  fitting procedure selects the template spectrum that minimizes the
  $\chi^2$, and therefore provides a value for each of stellar
  population properties: the \emph{age} of the best--fit model, the
  $e$-folding time of the SFR $\tau$, the extinction $A_{\rm V}$, and
  the stellar mass.  In addition, the instantaneous SFR is computed
  from the age and $\tau$.  During the fitting, only the observed
  bands corresponding up to the rest-frame $K_s$-band
  ($\lambda_{rest}< 2.5\mu$m) were used to avoid any dust--emission
  contamination.  While the details of the procedure and its
  uncertainties are extensively discussed in \citet{poz07}, we recall
  a few points here.  The median formal statistical uncertainties,
  derived from the width of the probability distributions in the
  fitting procedure, for each stellar population parameter (mass, age,
  SFR, $A_{\rm V}$, $\tau$) are approximately 10-20\%.  Using
  extensive Monte Carlo simulations, \citet{poz07} demonstrated that
  the overall internal accuracy of the measured stellar masses is
  $\sim 0.2$ dex.  In addition to this typical internal error,
  systematic errors should also be mentioned. One source of systematic
  error is the addition of secondary bursts to a continuous star
  formation history, which produces systematically higher (up to 40\%
  on average) stellar masses.  In addition, we note that the synthetic
  spectra including the TP-AGB phase \citep[][ Charlot \& Bruzual, in
  prep.]{mar05} produce, instead, a shift of $\sim 0.1$ dex towards
  lower stellar masses than models without this important phase
  \citep{bru03}, at least for our sample.  Finally, the uncertainty in
  the absolute value of the stellar mass, owing to assumptions about the
  IMF, is within a factor of two for the IMFs typically adopted in the
  literature.

The galaxies with $\log$(sSFR) = -7.9 are affected by the limit set to
be the minimum age of $9 \times 10^7$ yr.  Since an exponential star
formation history is assumed, the sSFR estimate is related to the age
in the following way: sSFR $\propto \exp{-t/\tau}$.

There are no strong correlations between the derived properties,
  apart from the following classical ones: (1) the age-metallicity
  correlation, and (2) the age-reddening correlation for red galaxies.
  The first relation is not applicable here because we do not derive the
  metallicity (it is fixed to be solar). The second one may be a concern
  as it can cause a degeneracy between dust-free galaxies with old
  stellar populations and dusty galaxies with young
  populations. However, SED fitting is able to break this degeneracy
  if sufficient bands are included around the (redshifted) 4000\AA\ break
  \citep[see for example][ for EROs, using the RJK bands]{poz00}.  In
  the case of $z=1.6$ galaxies, the 4000\AA\ break is shifted to
  1\,$\mu$m.  As our SED fitting of galaxies at $z=1.6$ includes bands
  up to 6.5\,$\mu$m, this degeneracy should not be a major concern.

\subsection{Comparison of field and overdensity samples}

To compare the properties of the member galaxies with those of field
galaxies, we selected a sample of 43 galaxies outside the peak, that
is 24 in the redshift interval $1.416 < z < 1.598$ and 19 in the
interval $1.624 < z < 1.840$, which we refer to as the field sample.
The field galaxies were selected from the GMASS sample and have
therefore similar information available, derived in an identical way,
to that for the spike galaxies.  Of the 43 field galaxies, 25 have
redshifts determined by GMASS only, and 8 have uncertain redshifts.
We further subdivide the 42 galaxies within the redshift peak into
those within the (spatially) \emph{high--density} region defined in
Sect.~\ref{sec:introduction} (21 galaxies), and those outside (also
21, \emph{low--density} region).

For an unbiased comparison of derived galaxy properties, it is
necessary to know whether there are any selection effects related to
the galaxy redshifts.  The mass limit is mainly defined by the m$_{\rm
  4.5\mu m} < 23.0$ limit, which sets a completeness limit of about
$M_* =3 \times 10^9 $M$_\odot$ at $z < 1.8$ (for the M05 templates
used). There are six galaxies with masses below this limit within
  and six outside the redshift spike.
We set the fitted mass of these galaxies to $\log(M_*) = 9.5
$M$_\odot$ when we compare the sample masses.
All SFRs below $\log($M$_\odot$ yr$^{-1})$ = -0.25 are set to this
lower limit because they are not detected in the $U'$ band and SFRs
below this value can therefore not be measured accurately. A lower
limit to the age (of $9 \times 10^7$ yr) has been set by the fitting
program.  This enforces an upper limit to the sSFR because the stellar
mass is computed by assuming an exponential star formation history
over the lifetime of the galaxy.  In Sect.\
  \ref{sec:spectroscopic_bias}, we demonstrated that the [\ion{O}{II}]
  detectability is not responsible for the high density of
  spectroscopically confirmed galaxies at $z=1.6$.  The contribution
  of [\ion{O}{II}] detected galaxies to the sample at $z>1.6$ is
  nevertheless much smaller, which might cause a bias in the sense
  that fewer galaxies with high star--formation rate are
  selected. However, the upper-left plot in Fig.\
  \ref{fig:gal_props_1} indicates that this is not the case. In
  contrast, the average SFR (derived from SED fitting) of
  spectroscopically confirmed $z>1.6$ galaxies is higher than in the
  spike.  We do not expect a redshift bias to be present for any of
the other properties measured.

We compare the following properties: stellar mass, star formation rate
(SFR), specific SFR (sSFR), i.e., SFR/mass, age, rest frame $B-I$
colour, spectroscopic galaxy class, and morphological class (see
Table~\ref{table:spike_gals}).  

The spectroscopic class is determined by the presence of absorption
lines only (1.0), emission lines only (2.0), or absorption lines with
some emission lines present (1.5), in the GMASS and ESO/GOODS spectra.
This classification corresponds roughly to passive, early--type (1.0),
actively star--forming, late--type (2.0), and intermediate--type (1.5)
galaxies.  We note that the spectral type correlates well with the
observed morphology, that is early--types have compact bulge-like
morphologies, while late--types often show evidence for disk--like or
irregular morphologies (see Figs.~\ref{fig:spectra1} to
\ref{fig:spectra4}).  There is also very good agreement between
the properties derived from the SED fitting and the spectra.
\citet{cim08} show that the galaxies in a sample of high--redshift
(supposedly) early--type galaxies from GMASS are red, have
spheroidal morphologies and spectra consistent with a population of
old, nearly passively evolving, stars.  In addition, the photometric
SED fitting of these galaxies gives consistent results.

Some properties of the spike and field samples are shown in
Fig.~\ref{fig:gal_props_1}, while the results of K-S tests, which
determine the probability that the properties of the various samples
are drawn from the same distribution, are listed in
Table~\ref{table:galprop}.  The values in this table are
probabilities, that is a lower number indicates a more significant
difference in values for the various samples.  The samples compared
are high versus low density, high versus field, low versus field, and
spike versus field, respectively.  The mean and the error in the mean
of these properties are listed for each sample in
Table~\ref{table:mm_galprop}.

\begin{table}
\caption{Mean and error in the mean of galaxy properties}
\label{table:mm_galprop}            
\centering                          
\begin{tabular}{llrrrrrr}
\hline\hline                        
    &    & High  & Low & Spike & Field \\
Prop&Unit& \multicolumn{4}{c}{Mean (Standard deviation)} \\
\hline
Mass & log M$_\odot$     & 10.2 (0.3) &  9.8 (0.2) & 10.0 (0.2)& 10.0 (0.2)\\
SFR  & log M$_\odot$ yr$^{-1}$\hspace{-4ex}
                         & 1.0 (0.2) &  1.4 (0.1) &  1.2 (0.1) &  1.5 (0.1)\\ 
sSFR & log yr$^{-1}$     & -9.2 (0.2) & -8.4 (0.1) & -8.8 (0.2) & -8.5 (0.1)\\
Age  & 10$^9$ yr         & 0.6 (0.1)  & 0.2 (0.1) & 0.4 (0.1)  & 0.3 (0.1) \\
$B-I$&                   & 0.9 (0.1)  & 0.7 (0.1) & 0.8 (0.1)  & 0.8 (0.1) \\
\hline
\end{tabular}
\end{table}

\begin{table}
\caption{Results of the K-S tests comparing galaxy properties}
\label{table:galprop}      
\centering                          
\begin{tabular}{lllll}
\hline\hline                 
Samples$^{\mathrm{a}}$
      &    High-Low &    High-Field &    Low-Field &   S-Field \\
Prop  &    K-S      &    K-S      &    K-S     &    K-S    \\
\hline                                             
Mass  &{\bf0.029}   &    0.092    &    0.462   &    0.843  \\
SFR   &    0.304    &    0.098    &    0.932   &    0.506  \\
sSFR  &{\bf0.004}   &{\bf0.010}   &    0.936   &    0.204  \\
Age   &{\bf0.029}   &{\bf0.018}   &    0.688   &    0.473  \\
$B-I$ &    0.071    &    0.462    &    0.418   &    0.871  \\
Class &    0.583    &{\bf0.023}   &    0.106   &    0.741  \\
Morph &    0.071    &{\bf0.011}   &    0.523   &    0.923  \\
\hline
\end{tabular}
\begin{list}{}{}
\item[$^{\mathrm{a}}$] High and low density region, in spike (S) and field (F)
\end{list}
\end{table}

\begin{figure} \centering 
  \includegraphics[width=\columnwidth]{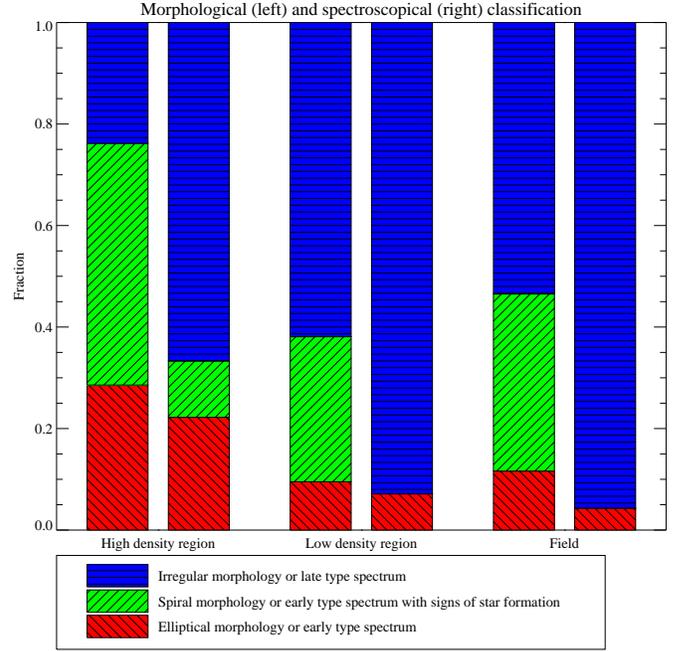}
  \caption{A comparison of the galaxy morphologies and spectral
    classes between the three samples (high and low density
    region, and field).}
  \label{fig:gal_props_hist}
\end{figure}

There are differences for all properties listed between the
high--density sample and the low--density or field sample, 
although some are only marginally significant.  We indicate the
probabilities of below 3\% in boldface in
Table~\ref{table:galprop}.  In particular, the differences in sSFR and
morphology are significant.  These are probably driven by the presence
of passively evolving galaxies with early type morphology in the
high--density region, which are not present in the low--density
region.

The mean differences between the high-density region and the field
amount to 0.2, 0.5, and 0.7 dex in mass, SFR, and sSFR, respectively,
and 0.3 Gyr in age.  Although there are several galaxies with masses
$M > 10^{10.5} M_\odot$ in the field sample, there are twice as many
galaxies with $M > 10^{10.7} M_\odot$ in the high--density sample as
in the field sample, which is twice as large (six against three). The
massive member galaxies are also those determined from their spectra
to be early types and have the lowest sSFRs.  Within the spike, there
are seven galaxies of early and intermediate spectral class, while
outside there is only one such galaxy.  We note that the more massive
galaxies have higher ages.

There is a notable difference in the angular distribution of
morphologically--classified elliptical galaxies in the spike: of the
eight elliptical galaxies, six are inside the high--density region
(see also Fig.~\ref{fig:geom}). Figure \ref{fig:gal_props_hist} shows
a comparison between the three samples of the morphologies and
spectral classes of the galaxies.

We also studied the correlation between the $z-J$ colour,
spectroscopic class, and morphology and the smaller--scale galaxy
density, as computed for and shown in Fig.~\ref{fig:geom}, for each
galaxy.  However, we did not find convincing evidence of any
significant correlations.

Another interesting comparison is listed in
Table~\ref{table:brgalprop}, where values below 0.001 are printed in
boldface.  We display the probabilities that the following samples
have properties derived from the same distribution: blue-member versus
red-member galaxies, blue-member versus blue-field galaxies,
red-member versus red-field galaxies, and blue-field versus red field
galaxies, respectively.  As expected, there are significant
differences between most properties (mass, sSFR, age, and morphology)
of the blue and red members.  The properties of blue galaxies inside
and outside the spike do not differ significantly, but the red members
inside the spike do have lower SFRs and sSFRs than the red field
galaxies (the mean sSFR having an order of magnitude difference).  In
the field, only the differences in mass and spectroscopic class are
significant between the blue and red galaxies.  This suggests that the
red galaxies inside the overdensity have evolved more rapidly than
those outside, while the blue galaxies appear to have had a similar
evolution inside and outside the overdensity.

\begin{table}
\caption{Blue and red galaxy property comparison}
\label{table:brgalprop}      
\centering                          
\begin{tabular}{lllll}
\hline\hline                 
Samples$^{\mathrm{a}}$
      &    BS-RS    &    BS-BF    &    RS-RF   &  BF-RF \\
Prop  &    K-S      &    K-S      &    K-S     &    K-S    \\
\hline                                             
Mass  &$<${\bf0.0001}  &    0.818    &    0.525   &{\bf0.0003}\\
SFR   &    0.0031  &    0.830    &    0.142   &    0.127  \\
sSFR  &$<${\bf0.0001}  &    0.592    &    0.103   &    0.451  \\
Age   &{\bf0.0002}  &    0.741    &    0.051   &    0.598  \\
Class &    0.0124  &    0.983    &    0.103   &    0.122  \\
Morph &{\bf0.0006}  &    0.980    &    0.421   &    0.200  \\
\hline
\end{tabular}
\begin{list}{}{}
\item[$^{\mathrm{a}}$] Blue (B) and red (R) galaxies, in 
spike (S) and field (F)
\end{list}
\end{table}

\begin{figure} \centering 
  \includegraphics[width=\columnwidth]{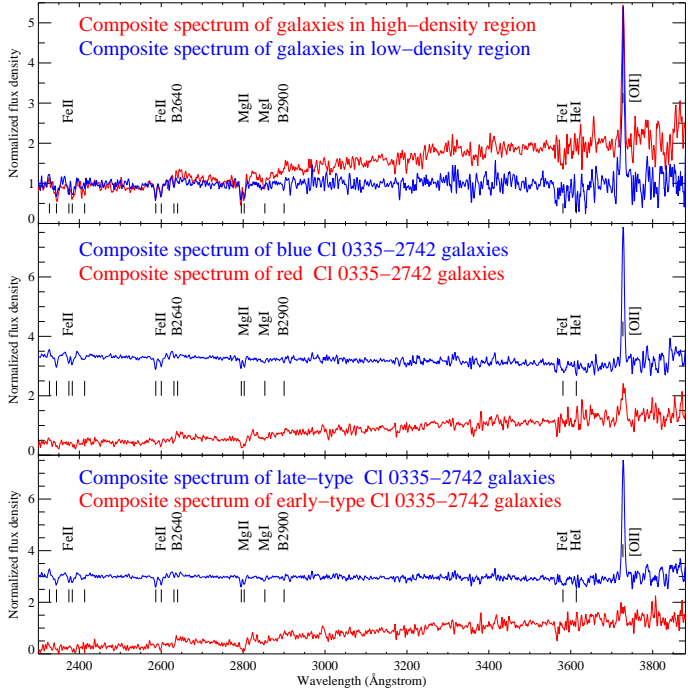}
  \caption{Shown here are rest--frame composite spectra of several
    groups of \cl\ galaxies with GMASS spectra.  The top panel shows a
    comparison of the composite spectra of galaxies in the
    high-density region (black) and in the low-density (grey). The
    middle panel shows a comparison of red (black) and blue (grey)
    \cl\ spectra, while the bottom panel shows the composite spectra
    of early (black) and late-type (grey) galaxies in \cl.  Absorption
    and emission features (possibly) identified are indicated by
    vertical lines below and above the spectra. In the middle and
    lower panel, the spectra are offset in vertical direction by an
    arbitrary amount to increase their visibility. }
  \label{fig:composite_spectra}
\end{figure}

\subsection{Composite spectra}

Finally, in Fig.~\ref{fig:composite_spectra}, we compare the
rest--frame composite spectra of the 18 \cl\ galaxies that have GMASS
spectra in the high-density region, with the composite spectrum of the
14 galaxies in the low-density region.  Also shown in this figure are
composite spectra of the 11 red and 21 blue galaxies and of the 7
early-- and 25 late--type galaxies in \cl.  The composites have been
normalized in the interval 2300 - 2700\,\AA, apart from the red and
early-type galaxy composites, which were normalized in the interval
3000 -- 3500\,\AA.  It is clearly noticeable that the composite
spectrum of the galaxies in the high-density region is far redder and
its [\ion{O}{II}] line equivalent width far lower (10\,\AA\ in
contrast to 40\,\AA) than the composite spectrum of the galaxies in
the low-density region.  The same is true for the red and blue, and
early-- and late--type galaxy composites (13\,\AA\ compared to 28\,\AA,
and 2\,\AA\ compared to 36\,\AA, respectively).

Apart from the 4000\,\AA\ and 3600\,\AA\ Balmer break (which are
  at $\lambda > 1$\,$\mu$m for $z>1.4$ galaxies), another strong
  age-dependent feature in the UV is a break in the region 2640 --
  2850\,\AA, which is produced by the combination of several strong
  absorption lines, including \ion{Mg}{II}$\lambda$2800.
  \citet{dad05} named this feature Mg$_{\rm UV}$ and defined it to be
  the ratio of (two times) the flux at $2625 < \lambda < 2725$\,\AA\
  to the flux in two 100\,\AA\ broad regions above and below this
  range.  \citeauthor{dad05} show that only stellar populations older
  than 0.1 (0.4) Gyr have Mg$_{\rm UV}$ significantly above 1.1 (1.4).
  Younger stellar populations and populations with constant star
  formation (but reddening E(B-V)=1.2) have Mg$_{\rm UV}$ near 1.0.
  This feature is, therefore, a key signature of passively evolving
  stellar populations because it is not present in young
  dust--reddened star--forming galaxies: this therefore allows us to
  break the age/dust degeneracy affecting red sources.  The
measurement of the Mg$_{\rm UV}$ index for our composite spectra
clearly differ: the high-density-region, red, galaxy composite has an
Mg$_{\rm UV}$ measurement of 1.2 in contrast to 1.1 for
low-density-region, blue galaxies, while the early- and late-type
composites have Mg$_{\rm UV}$ values of 1.4 and 1.1, respectively.

\section{Discussion}\label{sec:discussion}

We have presented evidence of a structure of galaxies at z = 1.6,
which may be a cluster in formation. The constituent galaxies exhibit
a galaxy overdensity of at least a factor of eight and a red-sequence of
passive galaxies, and they appear to be both more massive and evolved
than galaxies in the field at similar redshift.

We have compared this structure with clusters at $z > 1.0$, which have
similar properties, and often less reliably measured spectroscopic
redshifts available.  The differences in galaxy properties found by
ourselves between the high density region, the low density region,
and the field are consistent with the density relations found at $z
\sim 1$ by \citet{smi05} and \citet{pos05}, at $z = 1.24$ by
\citet{dem07}, and at $z = 1.5$ by \citet{fas08}: red early--types
dominate the cluster core, while blue late-types are found in the
outskirts of the cluster.

\subsection{Velocity dispersion}

The velocity dispersion determined for the galaxies in the
high--density region and for all galaxies in \cl\ are consistent with
a value of 500 km s$^{-1}$.  This value is about the average value
that \citet{hal04} and \citet{mil08} found for a sample of 26 clusters
at $0.40 < z < 0.96$ selected from the ESO Distant Cluster Survey
(EDisCS), but lower than the clusters at $0.8 < z < 1.3$ observed by
the ACS/HST in GTO time \citep{pos05}, most of which were first
detected by means of X-ray emission.  The spike at $z = 1.6$ is
therefore, as for the EDisCS clusters, more likely to be a progenitor of
\emph{typical} (less rich) low--redshift clusters \citep{mil08}.

The distribution of members of \cl\ in redshift space is bimodal
(Fig.~\ref{fig:zdistr_dz0.02}), with a main peak at $z=1.610$ and a
secondary peak at $z=1.602$.  \citet{dem07} found a similar bimodal
redshift distribution among the 38 confirmed members of the cluster
RDCS~J1252.9-2927 at $z = 1.237$.  In contrast to the latter cluster,
we do not find evidence that the substructure in redshift space
consists of two spatially separated groups, but it is probable that
also in the case of \cl\ the bimodality shows that the structure is
not yet virialized.

\subsection{Red sequence}

In the colour-magnitude diagram of the observed field (Fig.\
\ref{fig:cm_diagram}), we overplot a line that represents a
theoretical red sequence at $z = 1.60$, which corresponds to the
predicted magnitudes in $z$--band and $J$--band of the stellar
populations of elliptical galaxies that formed in a short time (0.5
Gyr) at $z_{\rm f} = 3.0$ \citep[provided by T.~Kodama, see][ based on
the code by \citealt{kod97}]{kod98}.  We indicate the (logarithm of)
stellar masses corresponding to the positions on the red sequence. The
early--type galaxy members are straddled about this red sequence,
with a scatter in $z-J$ of $0.147^{+0.063}_{-0.013}$ magnitude.  This
scatter also includes the scatter produced by photometric errors in
the measured colour. Subtracting this error (averaged in quadrature
over the seven galaxies) in quadrature, we obtain an intrinsic scatter
of 0.135 magnitudes.  The tightness in the observed red sequences at
low redshift is due to the large time difference between the epoch of
formation and the moment of observation.  Observing red sequences at
higher redshift, we approach this epoch of formation and expect the
scatter to increase due to small differences in the formation time and
evolution of the early--type galaxies.  \citet{sta98} found an
intrinsic scatter of about 0.05 for clusters from $z=0.1$ to $z=0.8$,
and \citet{ell06} measured 0.2$\pm$0.2 in a massive cluster at $z =
0.89$.  The scatter that we measure at $z = 1.61$ is consistent with
that expected from observations at lower redshift, taking into account
the younger average age of the stellar populations at this redshift.

\citet{del07} found a clear decrease in the ratio of luminous ($L >
0.4 L_*$) to faint red galaxies in clusters from $z \sim 0.8$ to $z
\sim 0.4$.  This increasing fraction of faint red galaxies towards
lower redshift could be explained if the RS of high redshift clusters
does not contain all of the progenitors of nearby RS cluster galaxies.
Instead, a significant fraction of these must have moved on to the RS
below $z \sim 0.8$.  Although Fig.~\ref{fig:cm_diagram} appears to
support this claim by the lack of galaxies with $z_{\rm phot} = 1.6$
on the red sequence at $J \gtrsim 23.0$, we note that, at these
fainter magnitudes, we have larger errors in the photometry and
photometric redshifts.  We are unable to confirm this decrement in the
number of faint galaxies on the RS, spectroscopically, because of the
limited completeness of our spectroscopy.

The red sequence appears to be ubiquitous in the field and in clusters
at low redshift, and is also detected at redshifts $z > 1$.
\citet{mcc07} detected one of the most distant red sequences
discovered so far, at $z = 1.5$, including at least ten galaxies with
very red optical--to--NIR and optical--to--MIR colours within a circle
of 128 kpc radius, one of which has a spectroscopic redshift of $z =
1.51$.  The total stellar mass of these galaxies is $\sim 8 \times
10^{11}$ M$_\odot$, which is similar to the total stellar mass
computed for \cl.  Six of the galaxies lie on a red sequence in a
$J-K$ CM-diagram, which is consistent with the colour-magnitude
relation for the Coma cluster evolved to $z=1.5$, assuming a
passively--evolving SSP with $z_f = 5$ \citep[according to][]{mcc07}.

The bimodality in galaxy colours up to $z \sim 2$ was also found by
\citet{cas08} in the GMASS field and \citet{gia05} in the HDF.  The
latter also showed that the blue and red galaxy populations most
probably had different evolutionary histories.  Using a much larger
sample, including $\sim$22000 galaxies selected over an area of 0.6
deg$^2$ from the Early Data Release of the UKIDSS Ultra Deep Survey,
\citet{cir07} found that the colour bimodality disappears at $z
\gtrsim 1.5$. Recently, \citet{wil09} showed, with a large
  $K$--selected sample of more than 30\,000 galaxies, that the bimodal
  distribution of star--forming and quiescent galaxies is still seen
  in a subsample of galaxies in the redshift range $1.5 < z < 2$, but
  not above $z=2$.  The clear bimodality demonstrated for \cl\ may
indicate that the colour bimodality develops earlier in high--density
regions than in a field sample, such as that employed by
\citeauthor{cir07} and \citeauthor{wil09}.

The relationship between galaxy colour and environment is studied in
more detail by \citet[][ CNC07]{coo07} employing a sample of 19\,464
galaxies selected from the DEEP2 Galaxy Redshift Survey.  They measured
the galaxy surface overdensity within redshifts bins up to $z \sim
1.4$ by calculating the distance to the third closest neighbour, and
the red fraction by considering ($U-B$)--restframe colour, which is
slightly dependent on luminosity.  The overdensities sampled range up
to about ten, which, as CNC07 state, does not cover the regime of
massive clusters.  CNC07 stated that the red fraction at $z>1.4$ is
not higher in high--density regions than in low--density regions, but
this conclusion is based mainly on the value of a single density bin,
for overdensities from 1.8 to 10. The large size of this bin may
  mask the higher fraction of red galaxies in the upper part of the
  bin.  Although, because of our much smaller galaxy sample, it is
  difficult for us to measure the overdensity in exactly the same way
  as CNC07, we obtain a mean overdensity of $>$8 for the galaxies
  in \cl.  The evidence of a red sequence in \cl\ seems therefore at
  variance with the conclusion by CNC07, but the degree of
  inconsistency is difficult to quantify.  The disagreement may be
  explained if we are sampling a regime higher in density
  than studied by CNC07.

\subsection{Dependence of galaxy properties on galaxy density}

We have investigated the dependence of galaxy properties on galaxy
density by comparing the ensemble properties of several samples
(spike, field, high, and low density) and by relating the properties
of individual $z=1.61$ galaxies to the measured local density.  The
latter method did not provide significant evidence of correlations
between the measured galaxy properties and the local galaxy density.
There may be several reasons for this lack of significant
correlations.  First, the local galaxy density is determined from the
secure spectroscopic members and the less secure photometric members
on a scale of $\sim$300 (physical) kpc, which should be optimal for
the detection of galaxy groups, but is rather small given the mean
distance to the closest neighbour among the 42 confirmed members in
the GMASS field, which is of the order of this scale.  We note that,
when comparing galaxy properties, only the spectroscopic members
were considered, although the photometric members have a significant
influence on the computation of the local galaxy density.  For
example, the strong density peak in the western part of the GMASS
field consists almost entirely of photometric members, which certainly
merit spectroscopic follow-up.  It is also possible that the influence
of galaxy density is notable only on larger scales, such as those of
the high density region, for which we do find a significant
correlation between density and all measured properties, most notably
morphology.

The environmental dependence of galaxy properties was studied at
higher redshift for a significant overdensity of galaxies at $z=2.300$
by \citet{ste05}. They found galaxies inside the overdensity to have
mean stellar masses that were higher and inferred ages that were
$\sim$2 times older than identically UV--selected galaxies outside the
structure.  This is similar to our results at $z=1.61$.  \citet{pet07}
augmented this analysis with one of $z=2.300$ galaxy morphologies but
did not find evidence of a difference between the proto-cluster
sample and a control (field) sample.  Very few of the $z=2.300$
galaxies appear to be regular ellipticals or spirals.  The
morphological difference between overdensity and field, which we found
at $z=1.61$, is due to the elliptical galaxies, which are also red
and, using the UV selection applied at $z=2.300$, could therefore not
have been found by \citet{ste05}.

The lower SFR that \cite{ste05} and ourselves found for galaxies in
the overdensities compared to those in the field appears to
contrast with the conclusion of \citet[][ EDB07]{elb07} that the
average SFR derived from FIR and UV photometry increases with galaxy
density in the redshift range $0.8\leq z \leq 1.2$.  However, the
range of projected galaxy densities probed by EDB07 is between 0.4 and
$\sim$3 Mpc$^{-2}$ in recessional velocity bins of $v = \sim1500$ km
s$^{-1}$.  At $z=1.6$, this corresponds almost exactly to the redshift
bin $1.600 < z < 1.623$, encompassing all 42 spike galaxies.  Using
this bin and the rather large boxes of 1.5$\times$1.5 Mpc$^{2}$ used
by EDB07, the projected galaxy density in the GMASS field is 5
Mpc$^{-2}$ on average and 10 Mpc$^{-2}$ for the highest density
regions.  Obviously, using smaller boxes for counting the galaxy
density, as also noted by EDB07, would result in even higher projected
galaxy densities (up to $\sim$40 Mpc$^{-2}$).  This shows that we are
probing higher densities than studied by EDB07 and their conclusions
do not apply to \cl.  Although this is probably not a statistically
significant result, Figs. 8 and 9 of EDB07 appear to indicate that the
SFR does indeed decrease at projected galaxy densities of $>$3
Mpc$^{-2}$.

\section{Summary and conclusions}\label{sec:summary}

  In this section, we summarize the results presented in this paper
  and add a concluding remark on the galaxy structure found.

\begin{itemize}

\item We carried out a spectroscopic survey, called GMASS, in a part
  of the CDFS, targeting galaxies detected with IRAC/Spitzer at
  4.5\,$\mu$m and pre--selected on the basis of a photometric redshift
  $z>1.4$. We obtain spectroscopic redshifts $z>1.4$ for 135 out of
  the $\sim$200 observed sources (Kurk et al., in prep.).

\item The redshift distribution of these sources shows a prominent
  peak at $z=1.6$ consisting of 42 galaxies.  We verified that the
  redshift distribution is not significantly affected by observational
  biases and find that this spike represents an overdensity in
  redshift space by a factor of at least eight. We call this structure
  \cl\ and the galaxies at $z=1.6$ member galaxies (of this
  structure).

\item The angular distribution of the member galaxies is not
  homogeneous: within the 7\arcmin$\times$7\arcmin\ GMASS field, there
  is a region of about 3\,Mpc$^2$ where the number density of members
  is five times higher than in the rest of the field.  We call this
  region the high density region.  The brightest member is located
  760\,kpc away from the centre of this region, within 65\,kpc of two
  other bright members.  These galaxies are not only the brightest but
  also the reddest members.

\item From the 42 redshifts of the members, we determine a velocity
  dispersion of $440^{+95}_{-60}$ km\,s$^{-1}$ for \cl\ and of
  $500^{+100}_{-100}$ km\,s$^{-1}$ for the 21 members in the high
  density region.  The redshift distribution of the 42 member galaxies
  is best fit by the sum of two Gaussian functions with a difference
  of $\delta z=0.008$ (or 920 km\,s$^{-1}$ in the reference frame of
  \cl) in their central redshifts. Although we do not find evidence of
  a spatial separation between the two groups of galaxies centered at
  $z=1.602$ and $z=1.610$, we cannot exclude the existence of
  substructure along the line of sight.

\item Using various methods to derive the mass which will be or is
  contained within \cl, we find a mass within the range $6 - 60 \times
  10^{13}$ M$_\odot$.  The irregularity in angular and redshift space,
  including at least two localized higher density peaks, suggests that
  the structure is not yet virialized but will evolve to become a galaxy
  cluster at a later cosmic time.

\item We used the 1\,Ms \emph{Chandra} observations to search of hot
  gas associated with the high density region but did not find evidence
  for an X-ray luminosity $> 3.5\times10^{43}$ erg\,s$^{-1}$.  This
  X--ray luminosity, however, is consistent with the expected X--ray
  emission for a cluster or group of galaxies with a velocity
  dispersion of 500 km\,s$^{-1}$.

\item In a colour--magnitude diagram, a bimodal distribution of member
  galaxies is seen with blue galaxies around $z-J\sim0.6$ and red
  galaxies, including all early--type members, at $z-J>1.3$.  The
  latter form a red sequence consistent with model galaxies composed
  of stellar populations formed at a redshift $z=3$.  The measured
  scatter around the red sequence of 0.135 magnitudes is consistent
  with the expected scatter at this redshift.  The bright tip of the
  red sequence is formed by the triplet of bright elliptical galaxies
  mentioned above.  We speculate that the clear bimodality seen in
  colour for \cl\ members, which is stronger than in field samples at
  this redshift, is due to the bimodality developing earlier in
  volumes of high galaxy density, consistent with galaxy formation
  models where massive galaxies form first in the strongest density
  peaks present in the early universe.

\item We compare various galaxy properties of the sample of member
  galaxies within the high density region with the sample of member
  galaxies outside this region and with a similarly sized \emph{field}
  sample of galaxies at spectroscopic redshifts determined by GMASS
  in the range $1.4 < z < 1.8$, but excluding the member galaxies at
  $z=1.6$. These galaxy properties include the galaxy type based on
  spectral features (either early or late), the galaxy morphology
  (either elliptical, spiral or irregular), broad band $B-I$ colour,
  and four properties derived from broad band SED fitting: mass, star
  formation rate, specific star formation rate and age.  The galaxies
  within the high density peak are on average older, more massive, and
  redder than those outside the redshift peak.  The SFR and sSFR of
  the members in the high density region are on average three and five
  times lower than in the field.  These results are consistent with
  current views on galaxy formation and evolution, where the most
  massive galaxies form preferentially in the highest density peaks of
  the cosmic galaxy distribution, which collapse first.  The massive
  galaxies that we observe in the overdensity also seem to evolve more
  rapidly than their lower mass counterparts: they have lower specific
  star formation rates, already at $z = 1.6$, which implies that they
  have formed most of their stars by this epoch.

\item We construct composite spectra of blue and red, early-- and
  late--type members and of members in the high and low density
  regions.  The composite spectra show features consistent with the
  results obtained above: the high density composite is clearly redder
  than the low density composite.  In addition, the [\ion{O}{II}] line
  equivalent width, a measure of star formation, is clearly lower in
  the high density composite than in the low density composite.

\end{itemize}

We note that this is the first and only structure of this kind known:
its redshift of $z=1.6$ is higher than that of any known galaxy
cluster and the structure contains spectroscopically--confirmed red,
early--type galaxies, whereas at higher redshifts, the known and
possibly only members of galaxy overdensities are blue,
star--forming galaxies are known (such as LBGs and LAEs).  \cl
may therefore represent a link between the $z<1.4$ galaxy clusters
and $z>2$ galaxy overdensities.  The extent of this structure is
not limited by the size ($\sim$7\arcmin$\times$7\arcmin) of the field
covered by spectroscopy carried out within the GMASS project and
neither does it seem to be limited by the size of the CDFS
\citep{cas07}.  Its extent of more than 10 Mpc therefore suggests it
is a sheet in the web--like distribution of galaxies.  The region with
the highest surface density of galaxies within this sheet, called \cl\
in this paper, already contains seven massive, passively evolving
galaxies and already has a velocity distribution typical of a large
group of galaxies in the local Universe.  Since the number of passive
galaxies, its velocity dispersion, and its mass can only increase
during the $\sim$10 Gyr between $z=1.6$ and $z=0$, reaching properties
typical of a present--day cluster of galaxies, we conclude that, by
observing \cl, we are witnessing the assembly of a cluster of
galaxies.

\appendix
\section{Spectra and images of the member galaxies}

\begin{figure*} 
\centering
  \includegraphics[width=\linewidth]{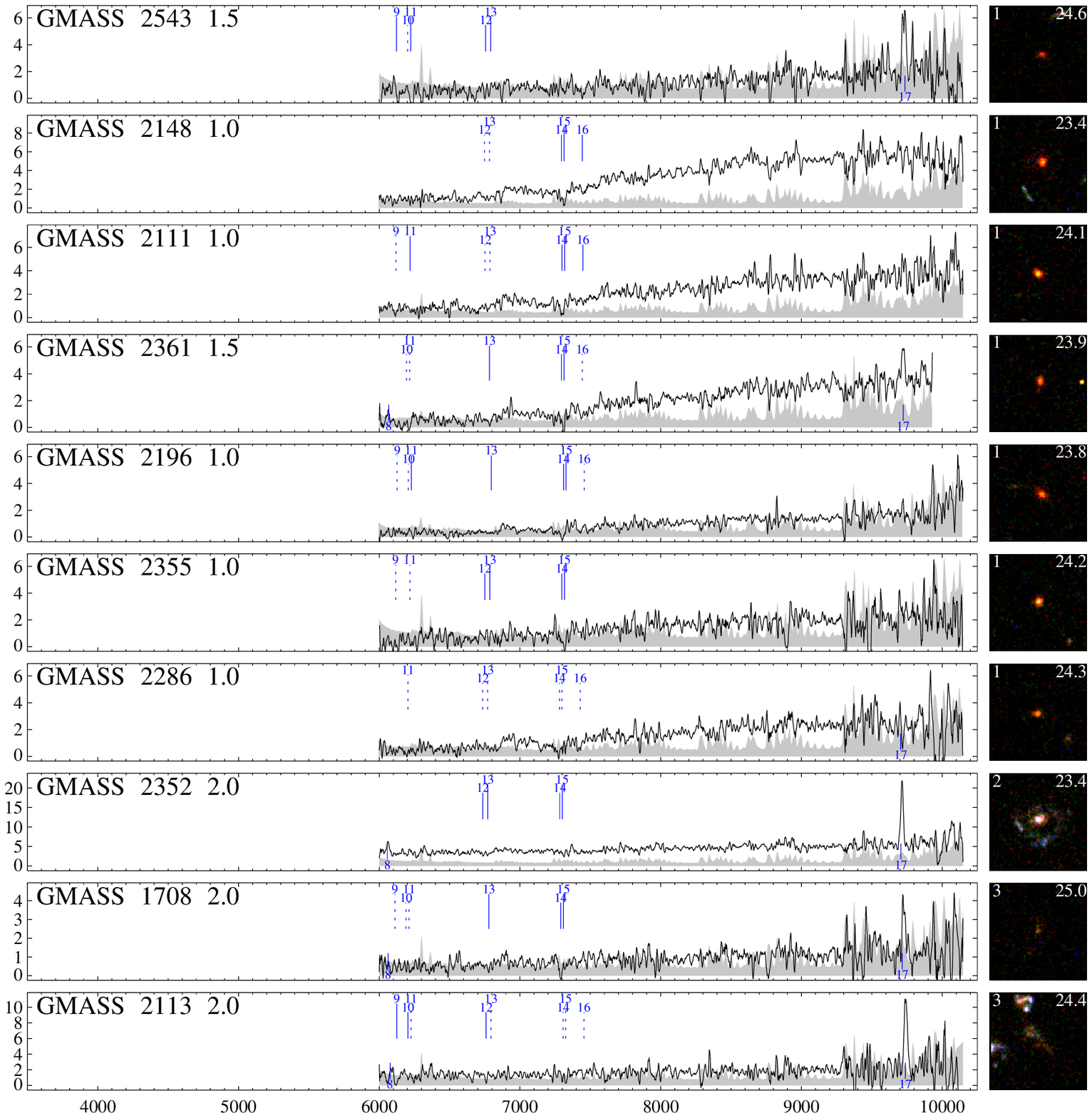}
  \caption{ Spectra and postage stamp images of the 42 member
    galaxies.  Wavelength in {\AA}ngstrom on the horizontal axis and
    flux in $10^{-19}$ erg\,s$^{-1}$\,\AA$^{-1}$\,cm$^{-2}$.
    Uncertainties caused by background noise are indicated by the
    underlying filled grey spectra.  As the flux calibration of the
    GOODS spectra is not very accurate \citep{van06} but the absolute
    flux level of these spectra seem systematically higher than the
    GMASS spectra (for the same magnitudes), they have been divided by
    a factor two.  The origin of the spectra is plotted: GMASS (Kurk
    et al., in prep.) or GOODS \citep{van06}, the GMASS identification
    numbers, and the spectroscopic classes (see text).  Robust
    (tentative) spectral features are shown by solid (dashed) lines
    with numbers (see below).  The postage stamps are constructed from
    the HST/ACS observations in the $B$, $V$, and $I$ bands, convolved
    with a Gaussian kernel.  Indicated are the morphological class (on
    the left) and the $z$ magnitude (on the right).  The spectra are
    sorted on $z-J$ colour, starting with the reddest galaxies.  The
    numbers for the spectral features refer to: (1)
    \ion{Si}{II}$\lambda$1527, (2) \ion{C}{IV}$\lambda$1550, (3)
    \ion{Fe}{II}$\lambda$1608, (4) \ion{Al}{II}$\lambda$1671, (5)
    \ion{Al}{III}$\lambda$1855, (6) \ion{C}{III}]$\lambda$1909, (7)
    \ion{Fe}{III}$\lambda$1926, (8) \ion{C}{II}]$\lambda$2326, (9)
    \ion{Fe}{II}$\lambda$2344, (10) \ion{Fe}{II}$\lambda$2375, (11)
    \ion{Fe}{II}$\lambda$2383, (12) \ion{Fe}{II}$\lambda$2587, (13)
    \ion{Fe}{II}$\lambda$2600, (14) \ion{Mg}{II}$\lambda$2796, (15)
    \ion{Mg}{II}$\lambda$2804, (16) \ion{Mg}{I}$\lambda$2853, (17)
    [\ion{O}{II}]$\lambda$3727.  }
  \label{fig:spectra1}
\end{figure*}

\begin{figure*} 
\centering
  \includegraphics[width=\linewidth]{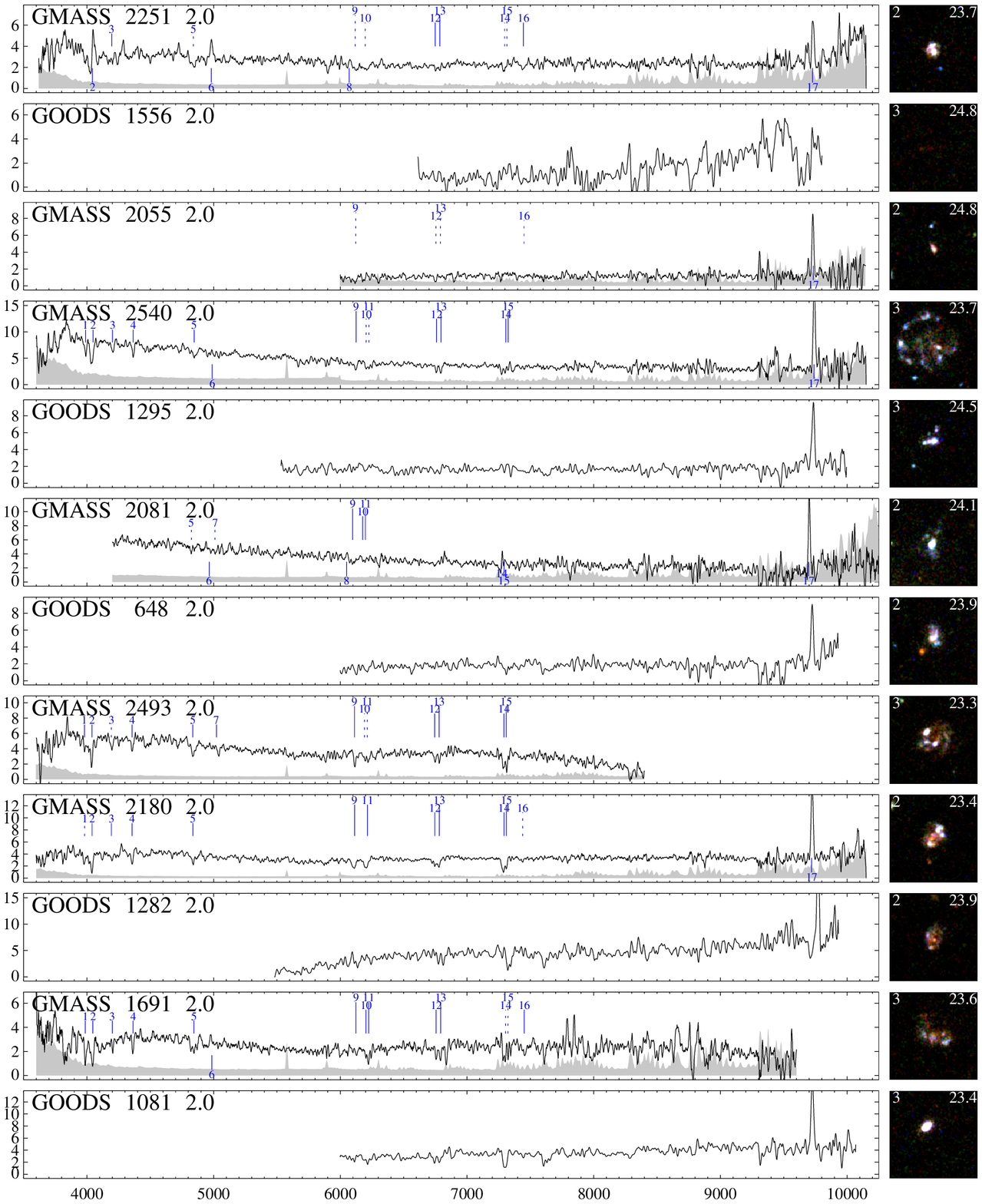}
  \caption{See Fig.~\ref{fig:spectra1} for description.}
  \label{fig:spectra2}
\end{figure*}

\begin{figure*} 
\centering
  \includegraphics[width=\linewidth]{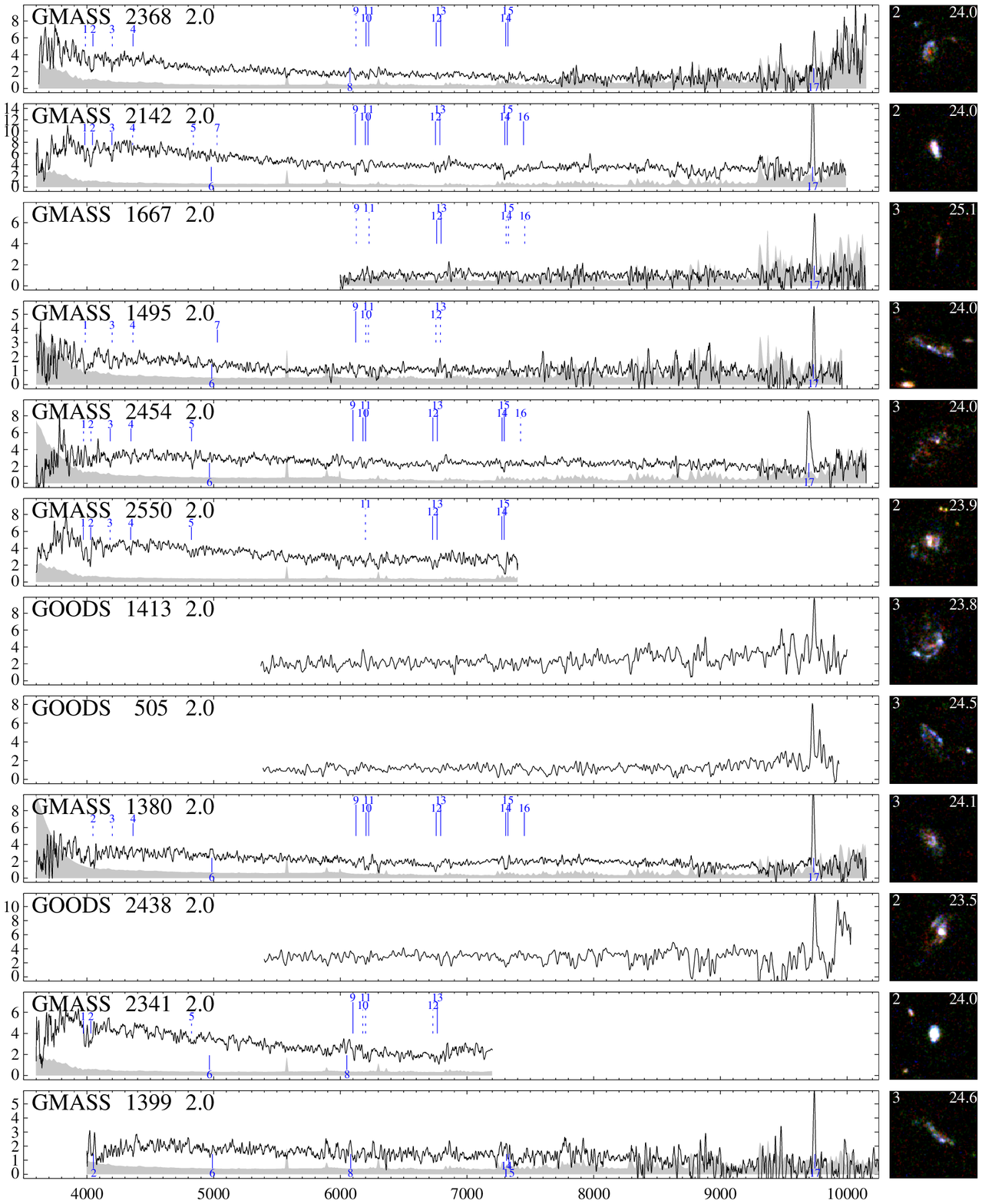}
  \caption{See Fig.~\ref{fig:spectra1} for description.}
  \label{fig:spectra3}
\end{figure*}

\begin{figure*} 
\centering
  \includegraphics[width=\linewidth]{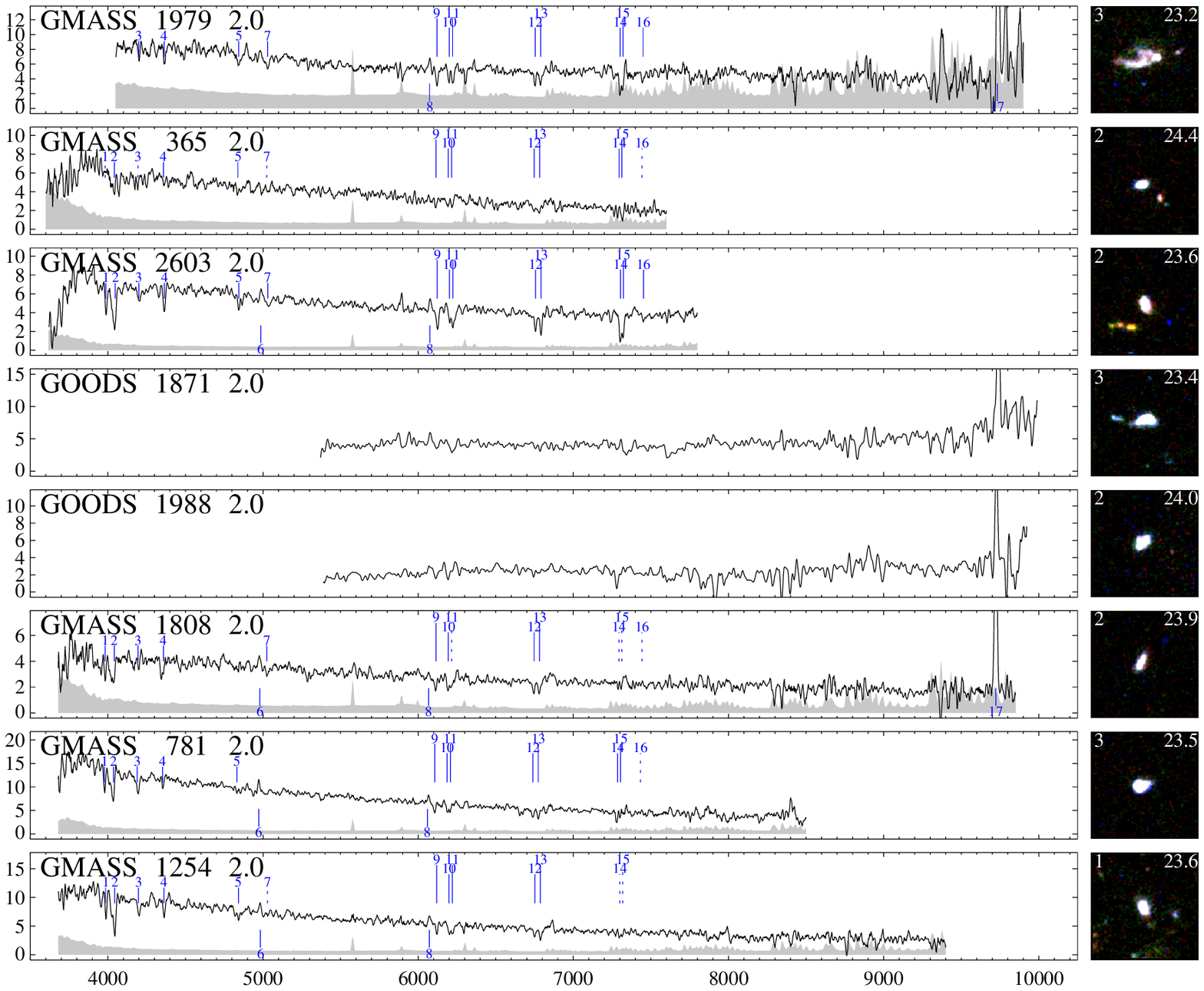}
  \caption{See Fig.~\ref{fig:spectra1} for description.}
  \label{fig:spectra4}
\end{figure*}

Figures \ref{fig:spectra1}, \ref{fig:spectra2}, \ref{fig:spectra3} and
\ref{fig:spectra4} show the spectra and images of the member galaxies.

\begin{acknowledgements}
  JDK is supported by the \emph{Deut\-sche
    For\-schungs\-ge\-mein\-schaft} (DFG), grant SFB-439.  This work
  is based on observations of the VLT Large Program 173.A-0687 carried
  out at the European Southern Observatory, Paranal, Chile.  This work
  is also based [in part] on observations made with the Spitzer Space
  Telescope, which is operated by the Jet Propulsion Laboratory,
  California Institute of Technology under a contract with NASA.
\end{acknowledgements}

\bibliographystyle{aa} 
\bibliography{kurk_gmass} 

\end{document}